\documentclass{article}

\usepackage{PRIMEarxiv}

\usepackage[utf8]{inputenc} 
\usepackage[T1]{fontenc}    
\usepackage{hyperref}       
\usepackage{url}            
\usepackage{booktabs}       
\usepackage{amsfonts}       
\usepackage{nicefrac}       
\usepackage{microtype}      
\usepackage{fancyhdr}       
\usepackage{graphicx}       

\usepackage{cite}
\usepackage{amsmath,amssymb}
\usepackage{amsthm}
\usepackage{tikz}
\usepackage{textcomp}
\usepackage{xcolor}

\theoremstyle{remark}\newtheorem{remark}{Remark}
\theoremstyle{plain}

\newtheorem{definition}{Definition}
\newtheorem{proposition}{Proposition}
\newtheorem{corollary}{Corollary}
\newtheorem{theorem}{Theorem}
\newtheorem{lemma}{Lemma}

\newcommand{\ie}{{\it i.e.,~}}

\newcommand{\LL}{\mathtt{Look}}
\newcommand{\CC}{\mathtt{Compute}}
\newcommand{\MM}{\mathtt{Move}}
\newcommand{\UU}{\mathtt{Update}}

\newcommand{\FSYNC}{\mathtt{FSYNC}}
\newcommand{\SSYNC}{\mathtt{SSYNC}}
\newcommand{\ASYNC}{\mathtt{ASYNC}}

\newcommand{\LUMI}{\mathcal{LUMI}}
\newcommand{\FSTA}{\mathcal{FSTA}}
\newcommand{\FCOM}{\mathcal{FCOM}}
\newcommand{\OBLOT}{\mathcal{OBLOT}}

\newcommand{\CABot}{\textsc{CABot}}
\newcommand{\Q}{\mathcal{Q}}
\newcommand{\Dir}{\mathcal{D}}
\newcommand{\rN}{\mathrm{N}}
\newcommand{\rS}{\mathrm{S}}
\newcommand{\rE}{\mathrm{E}}
\newcommand{\rW}{\mathrm{W}}
\newcommand{\stw}{\mathsf{w}}   
\newcommand{\ste}{\mathsf{e}}   
\newcommand{\sta}{\mathsf{a}}   
\newcommand{\str}{\mathsf{r}}   
\newcommand{\stl}{\mathsf{l}}   

\definecolor{cabW}{RGB}{190,200,255}
\definecolor{cabE}{RGB}{255,208,200}
\definecolor{cabA}{RGB}{228,140,216}
\definecolor{cabR}{RGB}{150,150,150}
\definecolor{cabL}{RGB}{110,195,110}
\definecolor{cabCa}{RGB}{214,236,196}
\definecolor{cabCb}{RGB}{160,210,125}
\definecolor{cabF}{RGB}{245,205,195}
\definecolor{cabWa}{RGB}{47,127,208}
\definecolor{cabWb}{RGB}{141,184,230}
\definecolor{cabWc}{RGB}{207,224,243}
\definecolor{cabWr}{RGB}{238,243,250}
\definecolor{cabT}{RGB}{232,72,63}
\definecolor{cabD}{RGB}{46,158,91}
\definecolor{cabWall}{RGB}{107,111,118}

\title{Autonomous Mobile Robots on a Computing Field}

\author{
  Yonghwan Kim \\
  Nagoya Institute of Technology \\
  Nagoya, Japan\\
  \texttt{kim@nitech.ac.jp} \\
   \And
  Gil-Tak Kong \\
  Busan National University \\
  Busan, Republic of Korea\\
  \texttt{ggt3080@gmail.com} \\
     \And
  Un-Sook Choi \\
  Tongmyong University \\
  Busan, Republic of Korea\\
  \texttt{choies@tu.ac.kr} \\
}

\begin{document}
\maketitle

\begin{abstract}
We introduce \CABot, a computational model for autonomous mobile robots in
which the discrete plane the robots operate on is not a passive stage but an
active computing medium. In the classical $\LL$-$\CC$-$\MM$ (\emph{LCM}) model
every increase in computational capability is obtained by strengthening the
robots themselves, with lights, memory or wider visibility, at a cost paid
once per robot. \CABot\ strengthens the environment instead: each cell of the
grid holds a state from a finite set $\Q$, the states evolve synchronously
under a uniform local rule that also senses whether a robot stands on the
cell, that is, the floor is a cellular automaton, and the $\LL$ operation
returns the states of the visible cells in addition to the positions of the
visible robots. The robots are as weak as we can make them: oblivious, silent,
lightless, myopic with visibility range~$1$, without multiplicity detection
and unable to write to the floor; they agree on the four directions and
operate fully synchronously. The number of field states $|\Q|$, rather
than any robot capability, becomes the measure of the resources required.
On an inert floor ($|\Q|=1$) such robots can neither form a line, nor gather
from scattered positions, nor explore a grid. We show that a handful of states
changes this. Line Formation from connected configurations is solvable with
four states, and \emph{field-silently}, so that the floor also reaches a fixed
point, with five. Gathering from arbitrary, possibly disconnected,
configurations is solvable with three states and Line Formation from arbitrary
configurations with six; we prove that neither can be solved field-silently
with any number of states. A single robot explores every finite grid forever
with four states and every torus with two, which is optimal. Further rule
sets, validated by simulation only, equalize the spacing of robots on a row,
fill a pattern painted on the floor and guide robots through a maze, and
indicate the range of the model.
\end{abstract}


\section{Introduction}
\subsection{Background}
A swarm of autonomous mobile robots is commonly studied
through the $\LL$-$\CC$-$\MM$~(\emph{LCM}) model~\cite{SY99,book19}.
Robots are points in the plane, anonymous and indistinguishable,
all executing the same algorithm (\ie homogeneous);
each repeatedly takes a snapshot of its surroundings ($\LL$),
computes a destination ($\CC$), and moves there ($\MM$).
The enduring appeal of the model lies in how little it assumes.
Robots have no identifiers, no memory that persists across cycles,
and no explicit means of communication.
Every positive result is therefore a statement about
how much coordination can be extracted from almost nothing.

The price of this minimalism is a long list of impossibility results~\cite{SY99,CFPS12}.
Symmetric configurations cannot be broken by identical deterministic robots;
oblivious robots cannot accumulate the history a task requires;
and myopic robots, which see only a bounded neighborhood,
cannot obtain the global view that many formations implicitly need.

The standard response has been to give the robots more.
Luminous robots ($\LUMI$) carry an externally visible light of constant size~\cite{DFPSY16};
the sub-models $\FSTA$ and $\FCOM$ separate its use as internal memory from
its use as communication~\cite{FSVY18};
other work enlarges the visibility radius or restricts the scheduler.
Together these form a well-understood hierarchy of models.
What they all share is that capability is added per robot:
a swarm of $n$ robots requires $n$ copies of the additional hardware,
and the robots themselves must become more complex, more expensive, and less disposable.

\subsection{Our Approach: Let the Floor Compute}

We take an orthogonal route.
Rather than asking what must be added to the robots,
we ask what must be added to the environment so that nothing has to be added to the robots.

Real robots never move in a vacuum.
Automated warehouse floors already carry fiducial markers and RFID grids
that vehicles read as they travel;
social insects coordinate not by direct communication but by depositing
and sensing traces in their surroundings,
a mechanism known as \emph{stigmergy}~\cite{TB99}.
In both cases the environment is part of the algorithm.
The LCM model, however, treats the plane as an inert set of coordinates.

We therefore promote the environment to a first-class computational entity
and call the resulting model \CABot\ (robots on a \emph{cellular automaton
field}). The grid on which the robots operate is itself a cellular automaton:
every cell carries a state from a finite set $\Q$,
and all cells update synchronously under a single uniform transition rule
whose input is the state of the cell, the states of its four neighbors, and
the presence or absence of a robot on the cell itself.
A robot's $\LL$~operation returns, in addition to the robot positions within its visibility range,
the states of the cells within that range. Informally: the floor is computing.

Three properties distinguish this from strengthening the robots.
\emph{Memory is attached to places, not to robots}: a light travels with its
carrier and vanishes when the robot leaves, whereas a field state stays where
it was produced, survives that departure, and can be read by a different robot
arriving later. \emph{Information propagates without being carried}: in the
plain LCM model information travels only as fast as the robots move or as far
as they see, while in \CABot\ it spreads one cell per step on its own,
which is what lets robots of visibility range~$1$ act on circumstances they
cannot see. \emph{The field is minimal infrastructure}: it is anonymous and
uniform, its rule is purely local, and it needs no identifiers, no broadcast
and no connectivity beyond nearest neighbors, so the question we ask is not
what an arbitrarily powerful environment can do but how little environmental
support suffices. Accordingly the robots are fixed at the weakest setting
throughout the paper, and the single resource we measure is the number
$|\Q|$ of field states.

\subsection{Field-Silence}

Once the environment computes, the usual notion of termination is too weak:
an algorithm may halt the robots in the desired configuration while the cells
beneath them keep changing forever, so the field still consumes energy and an
observer of the field cannot tell that the computation is over. We therefore
call an algorithm \emph{terminating} if every execution reaches a
configuration in which no robot moves again, and \emph{field-silent} if it
reaches one in which no cell changes state again either, by analogy with
silent self-stabilization~\cite{DGS99}. The distinction is not vacuous: our
four-state Line Formation algorithm is terminating but not field-silent, since
two cells at the ends of the formed line oscillate forever, and removing that
obstruction costs one additional state. For robots that start far apart the
distinction is even sharper: we show that no algorithm whatsoever can make
them meet field-silently.

\subsection{Our Contribution}
We formalize \CABot\ and study three fundamental problems in it,
\emph{Line Formation}, \emph{Gathering} and \emph{Exploration},
with robots that are as weak as the setting permits:
oblivious, silent, lightless, myopic with visibility range~$1$, without
multiplicity detection, agreeing on directions and operating fully
synchronously ($\FSYNC$). Table~\ref{tab:results} summarizes the results.

\begin{itemize}
\item \textbf{The model.} We define \CABot\ as an extension of LCM in which
$|\Q|$ measures the resources supplied by the environment, and we introduce
field-silence as a refinement of termination that is specific to models with
an active environment (Section~\ref{sec:model}). With $|\Q|=1$ the model is
the plain one, in which Line Formation from connected configurations,
Gathering and Line Formation from arbitrary configurations, and Exploration
are all unsolvable (Proposition~\ref{prop:inert}).
\item \textbf{Line Formation from connected configurations.} Four states
suffice on the infinite grid; the algorithm terminates but is not
field-silent (Section~\ref{sec:line}). Five states suffice for a field-silent
solution, and we identify the mechanism, a landing mark produced by decay
rather than by a persistent predicate, that makes the four-state algorithm
non-silent (Section~\ref{sec:silent}).
\item \textbf{Gathering from arbitrary configurations.} Robots that start on
arbitrary distinct cells, with no connectivity whatsoever, gather with three
states (Section~\ref{sec:gathering}). We also prove that no algorithm, with
any number of states, solves Gathering or Line Formation from arbitrary
configurations field-silently (Proposition~\ref{prop:nosilence}).
\item \textbf{Line Formation from arbitrary configurations.} Six states
suffice (Section~\ref{sec:linearb}). The algorithm combines the region
painting of the gathering algorithm with a \emph{landing zone}, a cell that
knows a robot is directly above it, which makes the approach collision-free
although the converging robots cannot see each other.
\item \textbf{Exploration.} A single robot visits every cell of every finite
torus and of every bounded grid infinitely often with four states, the floor
acting as a rotor-router; on tori two states suffice, which is optimal
(Section~\ref{sec:exploration}).
\item \textbf{Further examples.} Three more rule sets, validated by simulation
only, illustrate other uses of the floor: as a ruler that spaces robots
equally on a row, and as an excitable medium that calls robots to target cells
painted on the floor and routes them through a maze
(Section~\ref{sec:examples}).
\end{itemize}

The state counts are small enough to be striking:
four states amount to a cell being blank, occupied, or carrying a trace at one of two levels of freshness.
The computational capability that would otherwise demand a light or a memory on every robot
is recovered from a handful of states underfoot.

\begin{table}[t]
\centering
\caption{Summary of results. All robots are oblivious, silent, lightless and
myopic with visibility range~$1$, and operate under $\FSYNC$.}
\label{tab:results}
\small
\begin{tabular}{lllcll}
\toprule
Problem & Initial configuration & Arena & $|\Q|$ & Guarantee & Reference \\
\midrule
Line Formation & connected & $\mathbb{Z}^2$ & 4 & terminating & Theorem~\ref{thm:four} \\
Line Formation & connected & $\mathbb{Z}^2$ & 5 & field-silent & Theorem~\ref{thm:five}, Prop.~\ref{prop:silent} \\
Gathering & arbitrary & $\mathbb{Z}^2$ & 3 & terminating & Theorem~\ref{thm:gathering} \\
Line Formation & arbitrary & $\mathbb{Z}^2$ & 6 & terminating & Theorem~\ref{thm:linearb} \\
Gathering, Line Formation & arbitrary & $\mathbb{Z}^2$ & any & never field-silent & Prop.~\ref{prop:nosilence} \\
Exploration & one robot & torus, bounded grid & 4 & perpetual & Theorem~\ref{thm:explore4} \\
Exploration & one robot & torus & 2 & perpetual, optimal & Theorem~\ref{thm:explore2} \\
\midrule
each problem above & & & 1 & unsolvable & Prop.~\ref{prop:inert} \\
\bottomrule
\end{tabular}
\end{table}

\subsection{Organization}
Section~\ref{sec:related} surveys related work.
Section~\ref{sec:model} defines \CABot, the problems, and what an inert floor cannot do.
Sections~\ref{sec:line} and~\ref{sec:silent} present Line Formation from connected configurations with four and with five states.
Sections~\ref{sec:gathering} and~\ref{sec:linearb} treat Gathering and Line Formation from arbitrary configurations.
Section~\ref{sec:exploration} treats Exploration, Section~\ref{sec:examples} the further examples,
and Section~\ref{sec:conclusion} concludes with open problems.

\section{Related Work}
\label{sec:related}
\textbf{The LCM hierarchy.} The model is due to Suzuki and
Yamashita~\cite{SY99}, who established the role of symmetry in its
impossibility results; see~\cite{book19} for a survey. Because so much is
impossible for oblivious, silent, disoriented robots, later work adds
capability \emph{to the robots}: a persistent visible light
($\LUMI$~\cite{DFPSY16} and its sub-models $\FSTA$,
$\FCOM$~\cite{FSVY18}), axis agreement, multiplicity detection, or a
stricter scheduler. This is a hierarchy along one axis, how much each robot is
given. We add a second.

\textbf{Myopic robots on grids.} Gathering under limited visibility was
studied in~\cite{FPSW05}; on discrete spaces the radius is a sharp parameter,
since with radius~$1$ a robot sees only its four neighbors and connectivity
must be assumed. Gathering on grids has been studied with global
visibility~\cite{DDKN16} and with local visibility for connected
swarms~\cite{FJM17}, and arbitrary pattern formation on an infinite grid by
oblivious robots in~\cite{BAKS20}. With visibility~$1$ and no lights, a robot
that loses sight of all others can never move again; robots that start out of
sight of each other, as in Sections~\ref{sec:gathering}
and~\ref{sec:linearb}, are beyond the reach of the plain model altogether.

\textbf{Communicating through the environment.} Whiteboard
models~\cite{BFFS02} let an agent read and write bounded information at a
node, and pebble models~\cite{BFRSV98} let it drop markers; in both the
environment is \emph{passive}, nothing happening at a node unless an agent
stands on it. Ant-inspired covering algorithms let robots leave evaporating
traces~\cite{WLB99}. Stigmergy is active in our sense, pheromones diffusing and
evaporating while the ants are elsewhere~\cite{TB99}; it is a widespread
swarm-robotics heuristic but has, to our knowledge, no solvability theory.

\textbf{Rotor-routers.} In the rotor-router model~\cite{PDDK96}, also known as
the Propp machine or edge ant walk~\cite{YWB03}, every node holds a pointer
that cycles through its ports, and an agent leaves along the pointer, which
then advances. The walk is deterministic, covers every finite graph and locks
into an Eulerian circuit~\cite{YWB03,BGHIKK09,HLMPPW08}. In the literature the
pointer is part of the system by definition; in \CABot\ it is one particular
field, and Section~\ref{sec:exploration} shows that the floor can run it for an
oblivious robot that cannot write.

\textbf{Cellular automata and embedded agents.} A single agent reading and
writing the cells it visits is classical, Langton's ant being the canonical
example~\cite{Langton86}, but such systems are studied as dynamical systems
rather than as distributed algorithms. Excitable media, of which the
Greenberg--Hastings automaton~\cite{GH78} is the standard discrete model,
propagate waves that Section~\ref{sec:examples} uses as a signalling layer.
Programmable matter,
such as \emph{Amoebot}~\cite{DDGRSS14}, \emph{SILBOT}~\cite{GD20}, or \emph{Pairbot}~\cite{Kim25},
places computing entities on a lattice, but there
the particles \emph{are} the medium; in our model the field and the robots
are heterogeneous substrates and the \emph{empty} cells do most of the work.

\textbf{Silence.} Dolev et al.~\cite{DGS99} call a stabilizing algorithm
\emph{silent} when its communication registers eventually stop changing. In
the robot literature the analogue concerns the robots only, the environment
having nothing to change; once it computes the two separate, and
Sections~\ref{sec:line} and~\ref{sec:gathering} show the separation is real.

In short, memory resides in the environment rather than the robots, that
environment is active rather than passive, and the measured resource is
$|\Q|$ rather than a per-robot capability.

\section{The \CABot\ Model}
\label{sec:model}

We describe the two components of the system, the field and the robots,
then the cycle that couples them, and finally the problems.

\subsection{The Field}

The default arena is the infinite two-dimensional grid $\mathbb{Z}^2$, whose
elements we call \emph{cells}. We write
$\Dir=\{\rN,\rS,\rE,\rW\}$ for the four
cardinal directions and, for a cell $c=(x,y)$, we let
$c+\rN=(x,y+1)$, $c+\rS=(x,y-1)$, $c+\rE=(x+1,y)$ and
$c+\rW=(x-1,y)$. The neighborhood of $c$ is the von~Neumann
neighborhood $\mathcal{N}(c)=\{c+d \mid d\in\Dir\}$.

A \emph{cellular automaton field} (CA field) is a triple
$\mathcal{F}=(\Q,q_0,\delta)$ where $\Q$ is a finite set of
states, $q_0\in\Q$ is the \emph{quiescent} state, and
\[
  \delta:\ \Q\times\Q^{4}\times\{0,1\}\ \longrightarrow\ \Q
\]
is the transition function. A cell in state $q$ whose four neighbors are in
states $q_\rN,q_\rS,q_\rE,q_\rW$ and which does or
does not carry a robot, indicated by $b\in\{0,1\}$, moves to state
$\delta(q,q_\rN,q_\rS,q_\rE,q_\rW,b)$. All cells
apply $\delta$ simultaneously.

Three features matter. The field is \emph{anonymous and uniform}: one
$\delta$ governs every cell, cells have no identifiers or coordinates and see
only their four neighbors, so this is the weakest active infrastructure we
could place under the robots, not a powerful oracle. A cell detects a robot
\emph{only on itself}: the last argument of $\delta$ refers to $c$, never to
$\mathcal{N}(c)$, so a cell learns of a robot on a neighbor one round later,
through the state that neighbor takes in response; this delay is why several
algorithms below must distinguish a robot that has just arrived from one
standing still. Finally we require $\delta(q_0,q_0,q_0,q_0,q_0,0)=q_0$, so
that an untouched region stays quiescent and at every time only finitely many
cells are non-quiescent.

\paragraph{Finite arenas.} Exploration (Section~\ref{sec:exploration}) is
studied on finite arenas: the \emph{torus}
$T_{w,h}=\mathbb{Z}_w\times\mathbb{Z}_h$, in which $c+d$ is computed modulo
$(w,h)$, and the \emph{bounded grid}
$B_{w,h}=\{0,\dots,w-1\}\times\{0,\dots,h-1\}$, in which a cell on the border
lacks some neighbors. On a bounded grid a missing neighbor is read as
quiescent and unoccupied, by $\delta$ and by the robots alike, and a robot
that attempts to move to a missing cell stays where it is.

\subsection{The Robots}

A set of $n\ge 1$ robots occupies cells. They are
\emph{anonymous}, \emph{identical} and deterministic, \emph{oblivious},
\emph{silent} and \emph{lightless}, and \emph{myopic with visibility
range~$1$}. They agree on the four directions and on nothing else: no common
origin, no knowledge of $n$. Several robots may stand on the same cell; a cell
only senses whether it is occupied, and a robot only senses whether a
neighboring cell is occupied, so there is \emph{no multiplicity detection}.

Crucially, robots have \emph{no write operation}: every change in the field
is produced by $\delta$, which merely observes occupancy. A robot here is
exactly an $\OBLOT$ robot whose $\LL$~returns more, which distinguishes
\CABot\ from whiteboard and pebble models, where writing is an action of
the agent.

The \emph{view} of a robot standing on cell $c$ when the field assignment is
$\lambda:\mathbb{Z}^2\to\Q$ and the set of occupied cells is $P$ is
\[
  V(c)=\Bigl(\lambda(c);\ \bigl(\lambda(c+d)\bigr)_{d\in\Dir};\
             \bigl([\,c+d\in P\,]\bigr)_{d\in\Dir}\Bigr),
\]
an element of $\Q^{5}\times\{0,1\}^{4}$.
An \emph{algorithm} is a
function $\varphi:\Q^{5}\times\{0,1\}^{4}\to\Dir\cup\{\bot\}$,
$\bot$ meaning \emph{stay}. We abbreviate $\lambda_c=\lambda(c)$,
$\lambda_d=\lambda(c+d)$ and $\rho_d=[\,c+d\in P\,]$ for $d\in\Dir$,
and use the same abbreviations for the arguments of $\delta$.
A solution to a problem is a pair $(\varphi,\mathcal{F})$, and its cost is
$|\Q|$.

\subsection{The Cycle}

Time proceeds in rounds, and throughout this paper the system is fully
synchronous ($\FSYNC$): in every round all robots execute one
$\LL$-$\CC$-$\MM$ cycle in lockstep, and then the field performs one
$\UU$ step. A round thus consists of four phases, three carried out by the
robots and one by the field.
In round $t$, with occupied cells $P_t$ and assignment $\lambda_t$:
($\LL$) every robot on $c\in P_t$ obtains $V_t(c)$;
($\CC$) it evaluates $\varphi(V_t(c))$;
($\MM$) every robot with $\varphi(V_t(c))=d\ne\bot$ moves to $c+d$,
rigidly and by at most one cell, yielding $P_{t+1}$; and
($\UU$) every cell applies $\delta$ using the old states and its
\emph{new} occupancy,
$\lambda_{t+1}(c)=\delta(\lambda_t(c),(\lambda_t(c+d))_{d\in\Dir},[\,c\in P_{t+1}\,])$.

Phase $\UU$ reads $\lambda_t$ and writes $\lambda_{t+1}$, so the field update is
synchronous in the usual cellular-automaton sense. Phases $\LL$ and $\UU$ are the only
points of contact between the two layers: the robots read the field, and the
field reads the robots' positions.

A \emph{configuration} is $C_t=(P_t,\lambda_t)$ together with the number of
robots on each occupied cell, and the \emph{execution}
$C_0,C_1,\dots$ determined by $\varphi,\delta,C_0$ is unique since both are
deterministic and the scheduler is $\FSYNC$. Robots that share a cell have the
same view and therefore move together from then on. An algorithm is
\emph{collision-free} if $|P_t|=n$ always, that is, if no two robots ever
share a cell.

\paragraph{Other schedulers.} In the LCM model a scheduler decides which
robots are active and how their $\LL$, $\CC$ and $\MM$ phases interleave. In
\CABot\ a round has a fourth phase, $\UU$, which belongs to the field and
not to any robot. Under $\FSYNC$ there is one natural reading, the one above.
For semi-synchronous ($\SSYNC$) or asynchronous ($\ASYNC$) robots there is
not: one must decide whether the field advances in every round or only
around the robots that are active, how a field step is ordered relative to
the $\LL$ and $\MM$ phases of robots whose cycles overlap, and whether the
field and the robots share a clock at all. These choices are not equivalent,
and the timing on which our algorithms rely depends on them. We therefore
regard the definition of $\SSYNC$ and $\ASYNC$ schedulers as a part of the
model that remains to be designed (Section~\ref{sec:conclusion}), and every
definition, algorithm and result of this paper is stated for $\FSYNC$ only.

\subsection{Initial Configurations}

Initially $\lambda_0\equiv q_0$ and the robots occupy $n$ distinct cells. We
call such an initial configuration \emph{arbitrary}, and \emph{connected} if
in addition $P_0$ is connected under the neighborhood relation.
Connectivity is the standard assumption for myopic robots: one of visibility
range~$1$ that sees no one has nothing to act on. In \CABot\ it can be
dropped, because the field carries information between robots that cannot see
each other.

Where connectivity is assumed, it is assumed of $P_0$ only: our algorithms do
\emph{not} preserve it, and in most executions the swarm disconnects and later
re-forms. A myopic algorithm in the plain LCM model must maintain
connectivity, a robot losing sight of the swarm being lost to it; in \CABot\
the trace in the cells keeps a separated robot coupled to the rest, so the
visibility graph may break without harm.

\subsection{Termination and Field-Silence}

An execution $C_0,C_1,\dots$ is \emph{terminating} if $P_t=P_{t_0}$ for all
$t\ge t_0$ for some $t_0$, and \emph{field-silent} if $C_t=C_{t_0}$ for all
$t\ge t_0$, so that the whole system is a fixed point. An algorithm has one of
these properties if all its executions do. Field-silence implies termination;
the converse fails, as Sections~\ref{sec:line} and~\ref{sec:gathering} show.

\subsection{The Problems}

A set $P$ of cells is a \emph{line} if it consists of
$n$ consecutive cells of a single row or of a single column, that is, if
$P=\{(x_0+i,y_0)\mid 0\le i<n\}$ or $P=\{(x_0,y_0+i)\mid 0\le i<n\}$ for some
$(x_0,y_0)$.

\begin{definition}[Line Formation]
An algorithm $(\varphi,\mathcal{F})$ \emph{solves Line Formation from
connected (respectively arbitrary) configurations} if it is collision-free
and, for every $n\ge 1$ and every connected (respectively arbitrary) initial
configuration of $n$ robots, the execution is terminating and the set of
occupied cells it terminates in is a line. It solves the problem
\emph{field-silently} if in addition every execution is field-silent.
\end{definition}

\begin{definition}[Gathering]
An algorithm \emph{solves Gathering from connected (respectively arbitrary)
configurations} if, for every $n\ge 1$ and every connected (respectively
arbitrary) initial configuration of $n$ robots, the execution is terminating
and terminates with all robots on a single cell. It solves the problem
\emph{field-silently} if in addition every execution is field-silent.
\end{definition}

\begin{definition}[Exploration]
Let $G$ be a torus or a bounded grid. An algorithm \emph{solves Exploration
of $G$} if a single robot, started on any cell of $G$, visits every cell of
$G$ infinitely often. The \emph{cover time} is the number of rounds after
which every cell has been visited at least once.
\end{definition}

Line Formation forbids multiplicity and Gathering demands it; Exploration is a
non-terminating task, for which the relevant measure besides $|\Q|$ is the
cover time.

\subsection{What an Inert Floor Cannot Do}
\label{sec:inert}

With $|\Q|=1$ the field carries no information and \CABot\ is the plain model
of oblivious myopic robots on a grid. A view is then the set
$S\subseteq\Dir$ of directions in which a neighbor is occupied, and an
algorithm is one of the finitely many maps
$\varphi:2^{\Dir}\to\Dir\cup\{\bot\}$.

\begin{proposition}\label{prop:inert}
Let $|\Q|=1$.
\begin{itemize}
\item[(i)] No algorithm solves Line Formation from connected configurations.
\item[(ii)] No algorithm solves Gathering, or Line Formation, from arbitrary configurations.
\item[(iii)] No algorithm solves Exploration of a torus $T_{w,h}$ with $w,h\ge 2$, or of a bounded grid with at least two cells.
\end{itemize}
\end{proposition}
\begin{proof}
For (i) and (ii), termination with $n=1$ forces $\varphi(\emptyset)=\bot$, so
a robot with no neighbor never moves.

(ii) Two robots on non-adjacent cells both have the view $\emptyset$ forever
and never move, so they neither meet nor become adjacent.

(i) Only the eleven views with at most two occupied neighbors occur when
$n\le 3$, and $\varphi(\emptyset)=\bot$. The obstacle is easy to see for the
natural algorithms: if a robot with a single neighbor stays, which keeps the
two-robot lines in place, then in an L-shaped configuration of three robots
only the corner robot may move; if it stays the L-shape persists, and each of
its moves collides with an arm or leaves all three robots without a neighbor.
Other algorithms rotate a pair
before stopping it, and the complete case analysis is finite but tedious, so
we carried it out by exhaustive enumeration: of the $5^4$ assignments to the
four single-neighbor views, exactly nine make both connected configurations of
two robots terminate in a line without a collision, and for each of them
none of the $5^6$ assignments to the six two-neighbor views forms a line from
all six connected configurations of three robots. Hence every map fails on a
connected configuration of at most three robots.

(iii) A single robot has the same view in every round and therefore takes the
same action $a$ in every round. If $a=\bot$ it visits one cell. Otherwise, on
a torus it stays in one row or column and visits at most $\max(w,h)<wh$
cells; on a bounded grid it walks to the border and stays there forever.
\end{proof}

\begin{remark}\label{rem:connected-gathering}
Proposition~\ref{prop:inert} deliberately says nothing about Gathering from
\emph{connected} configurations. There an inert floor appears to suffice: the
field-free rule ``move east if the eastern or the northern neighbor is
occupied; otherwise move south if the southern neighbor is occupied; otherwise
stay'' gathers every connected configuration of at most $11$ robots (all
$185{,}416$ of them, checked exhaustively, each within $2(n-1)$ rounds) and
every random connected configuration of up to $800$ robots that we tried. This is
why we study Gathering from arbitrary configurations, where the field is
indispensable by Proposition~\ref{prop:inert}(ii).
\end{remark}

\section{Line Formation with Four States}
\label{sec:line}

\begin{figure}[t]
\centering
\footnotesize
\setlength{\tabcolsep}{2pt}
\begin{tabular}{@{}cccc@{}}
\begin{tikzpicture}[x=0.29cm,y=-0.29cm,baseline]
\fill[white] (-1,0) rectangle ++(1,1);
\draw[black!40,line width=0.2pt] (-1,0) rectangle ++(1,1);
\fill[white] (0,0) rectangle ++(1,1);
\draw[black!40,line width=0.2pt] (0,0) rectangle ++(1,1);
\fill[black] (0.5,0.5) circle (0.30);
\fill[white] (1,0) rectangle ++(1,1);
\draw[black!40,line width=0.2pt] (1,0) rectangle ++(1,1);
\fill[white] (2,0) rectangle ++(1,1);
\draw[black!40,line width=0.2pt] (2,0) rectangle ++(1,1);
\fill[white] (3,0) rectangle ++(1,1);
\draw[black!40,line width=0.2pt] (3,0) rectangle ++(1,1);
\fill[white] (-1,1) rectangle ++(1,1);
\draw[black!40,line width=0.2pt] (-1,1) rectangle ++(1,1);
\fill[white] (0,1) rectangle ++(1,1);
\draw[black!40,line width=0.2pt] (0,1) rectangle ++(1,1);
\fill[black] (0.5,1.5) circle (0.30);
\fill[white] (1,1) rectangle ++(1,1);
\draw[black!40,line width=0.2pt] (1,1) rectangle ++(1,1);
\fill[black] (1.5,1.5) circle (0.30);
\fill[white] (2,1) rectangle ++(1,1);
\draw[black!40,line width=0.2pt] (2,1) rectangle ++(1,1);
\fill[white] (3,1) rectangle ++(1,1);
\draw[black!40,line width=0.2pt] (3,1) rectangle ++(1,1);
\end{tikzpicture} & \begin{tikzpicture}[x=0.29cm,y=-0.29cm,baseline]
\fill[black!45] (-1,0) rectangle ++(1,1);
\draw[black!40,line width=0.2pt] (-1,0) rectangle ++(1,1);
\fill[white] (0,0) rectangle ++(1,1);
\draw[black!40,line width=0.2pt] (0,0) rectangle ++(1,1);
\fill[black] (0.5,0.5) circle (0.30);
\fill[black!45] (1,0) rectangle ++(1,1);
\draw[black!40,line width=0.2pt] (1,0) rectangle ++(1,1);
\fill[white] (2,0) rectangle ++(1,1);
\draw[black!40,line width=0.2pt] (2,0) rectangle ++(1,1);
\fill[white] (3,0) rectangle ++(1,1);
\draw[black!40,line width=0.2pt] (3,0) rectangle ++(1,1);
\fill[black!45] (-1,1) rectangle ++(1,1);
\draw[black!40,line width=0.2pt] (-1,1) rectangle ++(1,1);
\fill[white] (0,1) rectangle ++(1,1);
\draw[black!40,line width=0.2pt] (0,1) rectangle ++(1,1);
\fill[black] (0.5,1.5) circle (0.30);
\fill[white] (1,1) rectangle ++(1,1);
\draw[black!40,line width=0.2pt] (1,1) rectangle ++(1,1);
\fill[black] (1.5,1.5) circle (0.30);
\fill[black!45] (2,1) rectangle ++(1,1);
\draw[black!40,line width=0.2pt] (2,1) rectangle ++(1,1);
\fill[white] (3,1) rectangle ++(1,1);
\draw[black!40,line width=0.2pt] (3,1) rectangle ++(1,1);
\end{tikzpicture} & \begin{tikzpicture}[x=0.29cm,y=-0.29cm,baseline]
\fill[black!18] (-1,0) rectangle ++(1,1);
\draw[black!40,line width=0.2pt] (-1,0) rectangle ++(1,1);
\fill[black!45] (0,0) rectangle ++(1,1);
\draw[black!40,line width=0.2pt] (0,0) rectangle ++(1,1);
\fill[white] (1,0) rectangle ++(1,1);
\draw[black!40,line width=0.2pt] (1,0) rectangle ++(1,1);
\fill[black] (1.5,0.5) circle (0.30);
\fill[white] (2,0) rectangle ++(1,1);
\draw[black!40,line width=0.2pt] (2,0) rectangle ++(1,1);
\fill[white] (3,0) rectangle ++(1,1);
\draw[black!40,line width=0.2pt] (3,0) rectangle ++(1,1);
\fill[black!18] (-1,1) rectangle ++(1,1);
\draw[black!40,line width=0.2pt] (-1,1) rectangle ++(1,1);
\fill[white] (0,1) rectangle ++(1,1);
\draw[black!40,line width=0.2pt] (0,1) rectangle ++(1,1);
\fill[black] (0.5,1.5) circle (0.30);
\fill[white] (1,1) rectangle ++(1,1);
\draw[black!40,line width=0.2pt] (1,1) rectangle ++(1,1);
\fill[black] (1.5,1.5) circle (0.30);
\fill[black!18] (2,1) rectangle ++(1,1);
\draw[black!40,line width=0.2pt] (2,1) rectangle ++(1,1);
\fill[white] (3,1) rectangle ++(1,1);
\draw[black!40,line width=0.2pt] (3,1) rectangle ++(1,1);
\end{tikzpicture} & \begin{tikzpicture}[x=0.29cm,y=-0.29cm,baseline]
\fill[white] (-1,0) rectangle ++(1,1);
\draw[black!40,line width=0.2pt] (-1,0) rectangle ++(1,1);
\fill[black!45] (0,0) rectangle ++(1,1);
\draw[black!40,line width=0.2pt] (0,0) rectangle ++(1,1);
\fill[black!45] (1,0) rectangle ++(1,1);
\draw[black!40,line width=0.2pt] (1,0) rectangle ++(1,1);
\fill[white] (2,0) rectangle ++(1,1);
\draw[black!40,line width=0.2pt] (2,0) rectangle ++(1,1);
\fill[black] (2.5,0.5) circle (0.30);
\fill[white] (3,0) rectangle ++(1,1);
\draw[black!40,line width=0.2pt] (3,0) rectangle ++(1,1);
\fill[black!18] (-1,1) rectangle ++(1,1);
\draw[black!40,line width=0.2pt] (-1,1) rectangle ++(1,1);
\fill[white] (0,1) rectangle ++(1,1);
\draw[black!40,line width=0.2pt] (0,1) rectangle ++(1,1);
\fill[black] (0.5,1.5) circle (0.30);
\fill[white] (1,1) rectangle ++(1,1);
\draw[black!40,line width=0.2pt] (1,1) rectangle ++(1,1);
\fill[black] (1.5,1.5) circle (0.30);
\fill[black!18] (2,1) rectangle ++(1,1);
\draw[black!40,line width=0.2pt] (2,1) rectangle ++(1,1);
\fill[white] (3,1) rectangle ++(1,1);
\draw[black!40,line width=0.2pt] (3,1) rectangle ++(1,1);
\end{tikzpicture} \\
\begin{tabular}{@{}c@{}}$t{=}0$\\[-1pt]{\scriptsize initial}\end{tabular} & \begin{tabular}{@{}c@{}}$t{=}2$\end{tabular} & \begin{tabular}{@{}c@{}}$t{=}3$\\[-1pt]{\scriptsize slide (R1)}\end{tabular} & \begin{tabular}{@{}c@{}}$t{=}5$\end{tabular} \\[3pt]
\begin{tikzpicture}[x=0.29cm,y=-0.29cm,baseline]
\fill[white] (-1,0) rectangle ++(1,1);
\draw[black!40,line width=0.2pt] (-1,0) rectangle ++(1,1);
\fill[black!18] (0,0) rectangle ++(1,1);
\draw[black!40,line width=0.2pt] (0,0) rectangle ++(1,1);
\fill[black!18] (1,0) rectangle ++(1,1);
\draw[black!40,line width=0.2pt] (1,0) rectangle ++(1,1);
\fill[black!45] (2,0) rectangle ++(1,1);
\draw[black!40,line width=0.2pt] (2,0) rectangle ++(1,1);
\fill[black!45] (3,0) rectangle ++(1,1);
\draw[black!40,line width=0.2pt] (3,0) rectangle ++(1,1);
\fill[black!45] (-1,1) rectangle ++(1,1);
\draw[black!40,line width=0.2pt] (-1,1) rectangle ++(1,1);
\fill[white] (0,1) rectangle ++(1,1);
\draw[black!40,line width=0.2pt] (0,1) rectangle ++(1,1);
\fill[black] (0.5,1.5) circle (0.30);
\fill[white] (1,1) rectangle ++(1,1);
\draw[black!40,line width=0.2pt] (1,1) rectangle ++(1,1);
\fill[black] (1.5,1.5) circle (0.30);
\fill[white] (2,1) rectangle ++(1,1);
\draw[black!40,line width=0.2pt] (2,1) rectangle ++(1,1);
\fill[black] (2.5,1.5) circle (0.30);
\fill[white] (3,1) rectangle ++(1,1);
\draw[black!40,line width=0.2pt] (3,1) rectangle ++(1,1);
\end{tikzpicture} & \begin{tikzpicture}[x=0.29cm,y=-0.29cm,baseline]
\fill[white] (-1,0) rectangle ++(1,1);
\draw[black!40,line width=0.2pt] (-1,0) rectangle ++(1,1);
\fill[white] (0,0) rectangle ++(1,1);
\draw[black!40,line width=0.2pt] (0,0) rectangle ++(1,1);
\fill[white] (1,0) rectangle ++(1,1);
\draw[black!40,line width=0.2pt] (1,0) rectangle ++(1,1);
\fill[white] (2,0) rectangle ++(1,1);
\draw[black!40,line width=0.2pt] (2,0) rectangle ++(1,1);
\fill[white] (3,0) rectangle ++(1,1);
\draw[black!40,line width=0.2pt] (3,0) rectangle ++(1,1);
\fill[black!45] (-1,1) rectangle ++(1,1);
\draw[black!40,line width=0.2pt] (-1,1) rectangle ++(1,1);
\fill[white] (0,1) rectangle ++(1,1);
\draw[black!40,line width=0.2pt] (0,1) rectangle ++(1,1);
\fill[black] (0.5,1.5) circle (0.30);
\fill[white] (1,1) rectangle ++(1,1);
\draw[black!40,line width=0.2pt] (1,1) rectangle ++(1,1);
\fill[black] (1.5,1.5) circle (0.30);
\fill[white] (2,1) rectangle ++(1,1);
\draw[black!40,line width=0.2pt] (2,1) rectangle ++(1,1);
\fill[black] (2.5,1.5) circle (0.30);
\fill[black!18] (3,1) rectangle ++(1,1);
\draw[black!40,line width=0.2pt] (3,1) rectangle ++(1,1);
\end{tikzpicture} & \begin{tikzpicture}[x=0.29cm,y=-0.29cm,baseline]
\fill[white] (-1,0) rectangle ++(1,1);
\draw[black!40,line width=0.2pt] (-1,0) rectangle ++(1,1);
\fill[white] (0,0) rectangle ++(1,1);
\draw[black!40,line width=0.2pt] (0,0) rectangle ++(1,1);
\fill[white] (1,0) rectangle ++(1,1);
\draw[black!40,line width=0.2pt] (1,0) rectangle ++(1,1);
\fill[white] (2,0) rectangle ++(1,1);
\draw[black!40,line width=0.2pt] (2,0) rectangle ++(1,1);
\fill[white] (3,0) rectangle ++(1,1);
\draw[black!40,line width=0.2pt] (3,0) rectangle ++(1,1);
\fill[black!18] (-1,1) rectangle ++(1,1);
\draw[black!40,line width=0.2pt] (-1,1) rectangle ++(1,1);
\fill[white] (0,1) rectangle ++(1,1);
\draw[black!40,line width=0.2pt] (0,1) rectangle ++(1,1);
\fill[black] (0.5,1.5) circle (0.30);
\fill[white] (1,1) rectangle ++(1,1);
\draw[black!40,line width=0.2pt] (1,1) rectangle ++(1,1);
\fill[black] (1.5,1.5) circle (0.30);
\fill[white] (2,1) rectangle ++(1,1);
\draw[black!40,line width=0.2pt] (2,1) rectangle ++(1,1);
\fill[black] (2.5,1.5) circle (0.30);
\fill[black!45] (3,1) rectangle ++(1,1);
\draw[black!40,line width=0.2pt] (3,1) rectangle ++(1,1);
\end{tikzpicture} \\
\begin{tabular}{@{}c@{}}$t{=}6$\\[-1pt]{\scriptsize fall (R2)}\end{tabular} & \begin{tabular}{@{}c@{}}$t{=}8$\end{tabular} & \begin{tabular}{@{}c@{}}$t{=}9$\end{tabular}
\end{tabular}
\caption{An execution of the four-state algorithm, $n=3$. Discs are robots; shading
is the field state (white $q_0$, light $q_1$, dark $q_2$; a cell under a robot
is $q_3$). The top robot slides east at $t{=}3,5$ and falls at $t{=}6$. From
$t{=}8$ the robots never move again, yet the two cells at the ends of the line
keep swapping $q_2$ and $q_1$: terminating, not field-silent
(Prop.~\ref{prop:notsilent}).}
\label{fig:exec}
\end{figure}
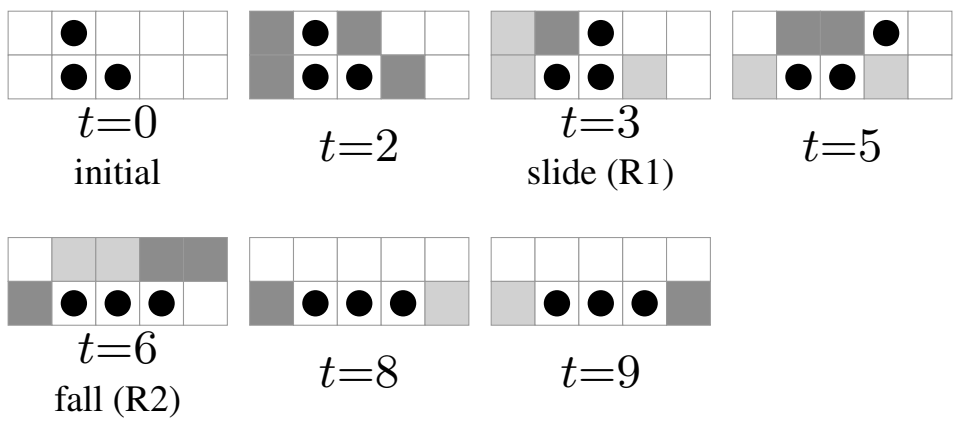

In this section and the next the initial configuration is connected.
Let $\Q=\{q_0,q_1,q_2,q_3\}$ with quiescent state $q_0$. Fix a
connected $P_0$ of $n$ robots and write
\[
  y_{\min}=\min\{y\mid (x,y)\in P_0\},\quad
  x_L=\min\{x\mid (x,y)\in P_0\}.
\]
Call row $y_{\min}$ the \emph{floor}. Robots fall towards it and slide east
along it, accumulating into one row; $q_3$ marks a robot and $q_2,q_1$ are a
trace of decreasing freshness left in the cells beside it (Fig.~\ref{fig:exec}).

\subsection{The Rules}

Define the decay map $\mathrm{dec}(q_3)=q_2$, $\mathrm{dec}(q_2)=q_1$,
$\mathrm{dec}(q_1)=\mathrm{dec}(q_0)=q_0$. The field transition $\delta$ is
given by the first applicable clause:

\begin{center}\small
\begin{tabular}{clc}
\hline
Rule & Condition on cell $c$ & Next state \\
\hline
F1 & $c$ carries a robot & $q_3$ \\
F2 & $(\lambda_c\in\{q_0,q_1\}) \wedge (\lambda_\rE=q_3 \vee \lambda_\rW=q_3)$ & $q_2$ \\
F3 & otherwise & $\mathrm{dec}(\lambda_c)$ \\
\hline
\end{tabular}
\end{center}

The robot algorithm is likewise the first applicable clause:

\begin{center}\small
\begin{tabular}{clc}
\hline
Rule & Condition & Action \\
\hline
R1 & $\neg\rho_\rN\wedge\rho_\rS\wedge\neg\rho_\rE\wedge\lambda_\rE=q_2\wedge\lambda_\rN=q_0$ & move E \\
R2 & $\neg\rho_\rS\wedge\lambda_\rS=q_1$ & move S \\
R3 & otherwise & stay \\
\hline
\end{tabular}
\end{center}

F2 is tested before F3, so a cell beside a robot is refreshed only from $q_0$
or $q_1$ and from $q_2$ it decays; the trace beside a stationary robot
therefore \emph{alternates}, which Lemma~\ref{lem:osc} turns into the timing
device R1 and R2 rely on.

\subsection{Basic Invariants}

\begin{lemma}[Marks are occupancy]\label{lem:mark}
For every $t\ge 1$ and every cell $c$, $\lambda_t(c)=q_3$ if and only if
$c\in P_t$.
\end{lemma}
\begin{proof}
F1 assigns $q_3$ exactly to occupied cells, and F2, F3 never produce $q_3$.
\end{proof}

\begin{lemma}[Trace oscillation]\label{lem:osc}
Let $c$ be unoccupied throughout rounds $t,t+1,\dots$ and suppose some cell of
$\{c+\rE,c+\rW\}$ is occupied throughout these rounds. Then for
every $t'\ge t+2$ we have $\lambda_{t'}(c)\in\{q_1,q_2\}$ and
$\lambda_{t'+1}(c)\ne\lambda_{t'}(c)$; that is, $\lambda(c)$ alternates
between $q_2$ and $q_1$ with period $2$ and never returns to $q_0$.
\end{lemma}
\begin{proof}
By Lemma~\ref{lem:mark} the occupied neighbor is at $q_3$ at every round
$t'\ge\max(t,1)$, and $\lambda_{t'}(c)\ne q_3$ at every such round. Hence at
every update from round $\max(t,1)$ on, F2 turns $\lambda(c)\in\{q_0,q_1\}$
into $q_2$, and F3 turns $\lambda(c)=q_2$ into $q_1$. So
$\lambda_{t'}(c)\in\{q_1,q_2\}$ for $t'\ge\max(t,1)+1$, and the two values
alternate.
\end{proof}

\begin{lemma}[The floor is frozen]\label{lem:floor}
For every $t$: (i) no robot occupies a cell of a row below $y_{\min}$;
(ii) every cell of row $y_{\min}-1$ is in state $q_0$; and (iii) no robot in
row $y_{\min}$ ever moves.
\end{lemma}
\begin{proof}
Simultaneous induction on $t$. Both (i) and (ii) hold at $t=0$ since
$\lambda_0\equiv q_0$. Assume them at $t$. For (iii), a robot at
$(x,y_{\min})$ cannot apply R2 because $\lambda_t(x,y_{\min}-1)=q_0\ne q_1$,
and cannot apply R1 because row $y_{\min}-1$ is unoccupied, so $\rho_\rS$
fails. Since R2 is the only rule that decreases a robot's row, and it is
unavailable in row $y_{\min}$, no robot enters row $y_{\min}-1$, giving (i) at
$t+1$. Then no cell of row $y_{\min}-1$ is occupied at round $t+1$; the
horizontal neighbors of such a cell lie in the same row and are in state
$q_0\ne q_3$, so F2 does not apply and F3 keeps $q_0$ at $q_0$; this gives
(ii) at $t+1$.
\end{proof}

\begin{lemma}[Collision-freedom]\label{lem:coll}
$|P_t|=n$ for every $t$; no two robots ever occupy the same cell.
\end{lemma}
\begin{proof}
A robot enters a cell $c$ only from $c+\rW$ by R1 or from
$c+\rN$ by R2, and both rules require $c$ to be unoccupied at the $\LL$
of that round, so no robot enters a cell that is occupied at the beginning of
the round. Two robots applying R1 in the same round have distinct sources and
hence distinct targets, and likewise for R2. The only remaining possibility is
that one robot applies R1 from $c+\rW$ while another applies R2 from
$c+\rN$. But R1 requires $\lambda(c)=q_2$ and R2 requires
$\lambda(c)=q_1$, and $q_1\ne q_2$.
\end{proof}

Collision-freedom is enforced neither by priorities nor by symmetry breaking,
but by the two movement rules reading \emph{different states of the same
cell}: $q_2,q_1$ are at once a landing signal and a mutual exclusion
mechanism. The same idea, in different guises, makes all our multiplicity-free
algorithms safe.

\subsection{Correctness}

\begin{lemma}[Columns]\label{lem:west}
(i) The lowest robot of a column never applies R1, so a column that contains
a robot contains one at every later round. (ii) At every round $t$ the columns
that contain a robot are $x_L,x_L+1,\dots,M_t$ for some $M_t\le x_L+n-1$.
In particular no robot ever occupies a cell with $x<x_L$.
\end{lemma}
\begin{proof}
R1 requires $\rho_\rS$, and the cell below the lowest robot of a column is
unoccupied. That robot can therefore only apply R2, which keeps it in its
column; this gives~(i). For~(ii), the columns that contain a robot of $P_0$
form an interval, because $P_0$ is connected, and its first column is $x_L$.
Robots move east and south only, so a robot enters a column only from the
column immediately to its west, and by~(i) no column is ever emptied. By
induction the columns that contain a robot form an interval that starts at
$x_L$ at every round; it has at most $n$ columns because there are $n$ robots.
\end{proof}

\begin{theorem}\label{thm:four}
The four-state algorithm solves Line Formation from connected configurations.
Moreover every execution terminates in
\[
  P_\infty=\bigl\{(x_L+i,\,y_{\min})\ \bigm|\ 0\le i<n\bigr\}.
\]
\end{theorem}

\begin{proof}[Proof sketch]
\emph{The robots stop.} Robots move east and south only. By
Lemma~\ref{lem:west} they stay in the columns $x_L,\dots,x_L+n-1$, and by
Lemma~\ref{lem:floor} in the rows from $y_{\min}$ up to the highest row of
$P_0$; there are at most $n$ such rows because $P_0$ is connected. Each robot
therefore moves at most $2(n-1)$ times, and from some round on the robots
stay forever on one set $S$ of cells.

\emph{Shape.} If all robots of $S$ are in row $y_{\min}$, then by
Lemma~\ref{lem:coll} they stand in $n$ distinct columns, which by
Lemma~\ref{lem:west} are $x_L,\dots,x_L+n-1$. So $S=P_\infty$.

\emph{No robot stops above the floor.} Suppose that the highest occupied row
$T$ of $S$ is above row $y_{\min}$. Once the robots are idle, every occupied
cell is in state $q_3$ (Lemma~\ref{lem:mark}), every unoccupied cell with an
occupied horizontal neighbor alternates between $q_2$ and $q_1$
(Lemma~\ref{lem:osc}), and every other cell reaches $q_0$ within three rounds
and stays there. Let $d$ be the cell of the easternmost robot of row $T$. Row
$T+1$ is empty, so $d+\rN$ ends in $q_0$, and $d+\rE$ alternates; hence if
$d+\rS$ were occupied, the robot on $d$ would apply R1. And if a robot of
row $T$ stood above an unoccupied cell with an occupied horizontal neighbor,
that cell would alternate and the robot would apply R2. Neither happens, so
the three cells $d+\rS+\rW$, $d+\rS$ and $d+\rS+\rE$ are unoccupied; then the
same holds for the robot on $d+\rW$, if there is one, and so on westwards.
The robots of row $T$ that form a contiguous run ending at $d$ thus
\emph{hover}: no robot stands below them, not even diagonally.

No such group exists in $P_0$, which is connected, and the trace is what keeps
the algorithm from creating one. For $s\ge 1$, if a cell $f$ or one of its
horizontal neighbors is occupied at round $s$ and $f$ is unoccupied at rounds
$s+1$ and $s+2$, then by F2 and F3 $\lambda(f)=q_1$ at round $s+1$ or at round
$s+2$, and a robot that stands above $f$ at that round falls. A robot whose
support moves away, or which slides off its support, is thus pulled down
behind it. Turning this into a proof that no hovering group ever arises
requires a case analysis of the relative timing of the last moves around the
group, which we omit. Instead, we verified the theorem exhaustively for all
$691{,}277$ connected configurations of at most $12$ robots
(Section~\ref{sec:sim}).
\end{proof}

\begin{corollary}\label{cor:rounds}
In every execution the robots stop within $O(n^2)$ rounds.
\end{corollary}
\begin{proof}
By the first step of the proof of Theorem~\ref{thm:four} the robots make at
most $2n(n-1)$ moves in total. Suppose that no robot moves in rounds
$t,\dots,t+3$, so that $P_t=\dots=P_{t+4}$. As in Lemmas~\ref{lem:mark}
and~\ref{lem:osc}, at each of the rounds $t+2$, $t+3$ and $t+4$ every occupied
cell is in state $q_3$, every unoccupied cell with an occupied horizontal
neighbor is in state $q_1$ or $q_2$ and changes at each update, and every
other cell is in state $q_0$. Hence $C_{t+4}=C_{t+2}$, the execution is
periodic from round $t+2$ on, and no robot ever moves again. So at most three
idle rounds lie between two consecutive moves, and the last move occurs
before round $8n(n-1)$.
\end{proof}

\subsection{The Algorithm Is Not Field-Silent}

\begin{proposition}\label{prop:notsilent}
The four-state algorithm is terminating but not field-silent. In the terminal
configuration exactly two cells, namely $(x_L-1,y_{\min})$ and
$(x_L+n,y_{\min})$, change state forever, alternating between $q_2$ and $q_1$
with period $2$; every other cell is fixed.
\end{proposition}
\begin{proof}
Termination is Theorem~\ref{thm:four}. In $P_\infty$ every occupied cell is
permanently $q_3$ by Lemma~\ref{lem:mark}. An unoccupied cell $c$ is refreshed
by F2 only if one of $c+\rE$, $c+\rW$ is occupied, that is, only if
$c$ lies in row $y_{\min}$ and is horizontally adjacent to the line; the only
such cells are $(x_L-1,y_{\min})$ and $(x_L+n,y_{\min})$, and by
Lemma~\ref{lem:osc} both alternate between $q_2$ and $q_1$ forever. Every
other unoccupied cell has no horizontal neighbor in state $q_3$, so F3 applies
at every round and it reaches $q_0$ within three rounds and stays there.
\end{proof}

The obstruction is structural. The landing signal that R2 waits for is
produced by \emph{decay}: a cell must fall from $q_2$ to $q_1$ for a robot
above it to move in. As long as a robot stands beside that cell, F2 keeps
pushing it back up to $q_2$, and the two rules chase each other forever. The
next section shows that separating the two roles the trace plays, a persistent
``you may land here'' predicate and a transient timing signal, removes the
oscillation at the cost of a single extra state.

\section{Field-Silent Line Formation with Five States}
\label{sec:silent}

By Proposition~\ref{prop:notsilent} the pair $q_2,q_1$ is asked to do two
incompatible jobs: to be a \emph{persistent} predicate marking a legal
landing site for as long as it is one, and a \emph{transient} signal, since
R2 waits for the decay $q_2\to q_1$ while R1 reads $q_2$ and that phase
difference is what keeps the rules from colliding. A persistent predicate
beside a stationary robot cannot decay; a transient one must. We separate the
jobs: the landing predicate becomes unconditional, and the timing moves onto
the robots' own cells. One extra state pays for both.

\subsection{The Rules}

Let $\Q=\{q_0,\ell_1,\ell_2,m^-,m^+\}$ and $M=\{m^-,m^+\}$. A cell
in $M$ carries a robot, $m^-$ if it arrived last round and $m^+$ otherwise;
$\ell_2,\ell_1$ say the cell is a legal landing site, $\ell_2$ while the
reason holds and $\ell_1$ for one round after. The field transition is:

\begin{center}\small
\begin{tabular}{clc}
\hline
Rule & Condition on cell $c$ & Next state \\
\hline
G1 & $c$ carries a robot and $\lambda_c\in M$ & $m^+$ \\
G2 & $c$ carries a robot & $m^-$ \\
G3 & $\lambda_\rE\in M$ or $\lambda_\rW\in M$ & $\ell_2$ \\
G4 & $\lambda_c\in M$ & $\ell_2$ \\
G5 & $\lambda_c=\ell_2$ & $\ell_1$ \\
G6 & otherwise & $q_0$ \\
\hline
\end{tabular}
\end{center}

\begin{center}\small
\begin{tabular}{clc}
\hline
Rule & Condition & Action \\
\hline
S1 & $\neg\rho_\rN\wedge\rho_\rS\wedge\neg\rho_\rE\wedge\lambda_\rE=\ell_2\wedge\lambda_\rN=q_0$ & move E \\
S2 & $\neg\rho_\rS\wedge\lambda_\rS\in\{\ell_1,\ell_2\}\wedge\lambda_c=m^+$ & move S \\
S3 & otherwise & stay \\
\hline
\end{tabular}
\end{center}

Three changes matter. \emph{The landing predicate is unconditional}: G3
fires whatever the cell's state, whereas F2 refreshed only from $q_0,q_1$, so
a cell beside a motionless robot sits at $\ell_2$ and stays there, which alone
removes the oscillation. \emph{The wake is kept, and fades}: G4 turns a cell
a robot has just left into $\ell_2$ and G5, G6 let it fade, which is needed
because a robot whose support descends would otherwise be stranded with no
landing site below and, by S1, no support to slide on; being
self-extinguishing, the wake is harmless for silence. \emph{The robot mark is
split}: S2 requires $\lambda(c)=m^+$, so a robot waits one round after
arriving, which replaces the $q_2/q_1$ phase difference as the
collision-avoidance mechanism, and $\ell_1$ exists so that a landing site does
not vanish during that wait.

\subsection{Correctness}

Lemma~\ref{lem:floor} (the floor is frozen) and Lemma~\ref{lem:west}
(columns) carry over verbatim: their proofs use only that a row with no
robots stays at $q_0$, that robots move east and south, and that the eastward
rule requires a robot below. The remaining invariants take the following
form.

\begin{lemma}\label{lem:mark5}
For every $t\ge 1$ and every cell $c$: $\lambda_t(c)\in M$ if and only if
$c\in P_t$. Moreover $\lambda_1(c)=m^-$ for every $c\in P_1$, and for
$t\ge 2$, $\lambda_t(c)=m^+$ if and only if $c\in P_t\cap P_{t-1}$.
Finally, if $c\in P_t\cap P_{t-1}$ then the same robot occupies $c$ at both
rounds.
\end{lemma}
\begin{proof}
Only G1 and G2 produce states of $M$, and they apply exactly to occupied
cells, which gives the first claim. G1 additionally requires
$\lambda_{t-1}(c)\in M$, which fails for $t=1$ because $\lambda_0\equiv q_0$
and is equivalent to $c\in P_{t-1}$ for $t\ge 2$; this gives the second. For
the last claim, a robot enters $c$ only by S1 or S2, and both require $c$ to
be unoccupied at the $\LL$ of that round, so no robot can enter $c$ in a round
in which another robot leaves it.
\end{proof}

So $\lambda(c)=m^+$ reads to the robot on $c$ as ``I did not move last
round'': the only history in the system, and it sits in the floor.

\begin{lemma}[The landing predicate is stable]\label{lem:stable}
Let $c$ be unoccupied at rounds $t,t+1,\dots$ and let one of
$c+\rE,c+\rW$ be occupied at all these rounds. Then
$\lambda_{t'}(c)=\ell_2$ for every $t'\ge t+2$.
\end{lemma}
\begin{proof}
By Lemma~\ref{lem:mark5} the neighbor is in $M$ at every round
$\ge\max(t,1)$, so G3 applies to $c$ at every update from then on, whatever
$\lambda(c)$ is.
\end{proof}

\begin{lemma}[Collision-freedom]\label{lem:coll5}
$|P_t|=n$ for every $t$.
\end{lemma}
\begin{proof}
As in Lemma~\ref{lem:coll}, the only dangerous case is a robot applying S1
from $b=c+\rW$ and a robot applying S2 from $a=c+\rN$ in the same
round $t$. Nobody moves in round $0$, since S1 needs $\ell_2$ and S2 needs
$m^+$; and in round $1$ every robot cell is $m^-$ by Lemma~\ref{lem:mark5}, so
S2 fails. Let $t\ge 2$ and note $a=(b+\rN)+\rE$.

If $a\in P_{t-1}$, then $\lambda_{t-1}(a)\in M$ by Lemma~\ref{lem:mark5}. If
$b+\rN$ is occupied at round $t$, S1 fails at $b$ because of $\rho_\rN$.
Otherwise G3 applied to $b+\rN$ at the update ending round $t-1$, so
$\lambda_t(b+\rN)=\ell_2\neq q_0$ and S1 fails at $b$ as well.

If $a\notin P_{t-1}$, then $a\in P_t\setminus P_{t-1}$, so
$\lambda_t(a)=m^-\neq m^+$ by Lemma~\ref{lem:mark5}; hence S2 fails at $a$.

The two cases are exhaustive, so at most one of the two robots moves. Two
robots applying the same rule have distinct sources and hence distinct
targets, and neither rule enters an occupied cell.
\end{proof}

The model's one-round delay is exactly what would let both robots move;
splitting the robot mark converts it into a one-round wait on the faller.

\begin{theorem}\label{thm:five}
The five-state algorithm solves Line Formation from connected configurations,
and every execution terminates
in the configuration $P_\infty=\{(x_L+i,y_{\min})\mid 0\le i<n\}$ of
Theorem~\ref{thm:four}, within $O(n^2)$ rounds.
\end{theorem}
\begin{proof}[Proof sketch]
The argument of Theorem~\ref{thm:four} applies with Lemmas~\ref{lem:stable}
and~\ref{lem:coll5} in place of Lemmas~\ref{lem:osc} and~\ref{lem:coll}. The
robots make at most $2n(n-1)$ moves, and if they stop with every robot in row
$y_{\min}$ they stop in $P_\infty$, for the same reasons. If no robot moves in
rounds $t,t+1,t+2$, then at round $t+2$ every occupied cell is $m^+$
(Lemma~\ref{lem:mark5}), every unoccupied cell with an occupied horizontal
neighbor is $\ell_2$ (Lemma~\ref{lem:stable}) and every other cell is $q_0$,
so $C_{t+3}=C_{t+2}$ and no robot ever moves again; this gives the bound of
$O(n^2)$ rounds. In such a configuration the easternmost robot of the highest
row applies S1 if a robot stands below it, and a robot above an unoccupied
cell with an occupied horizontal neighbor applies S2. Progress is easier here
than with four states, as no robot waits for a phase and a landing site stays
available once it exists. As before, robots could stop above the floor only
in a hovering group, and the wake of G4 plays the role of the trace; we again
omit the case analysis and verified the theorem exhaustively for all connected
configurations of at most $12$ robots (Section~\ref{sec:sim}).
\end{proof}

\subsection{Field-Silence}

\begin{proposition}\label{prop:silent}
The five-state algorithm is field-silent. Three rounds after the robots reach
$P_\infty$ the field is constant, with $\lambda(c)=m^+$ on the $n$ cells of
the line, $\lambda(c)=\ell_2$ on the two cells $(x_L-1,y_{\min})$ and
$(x_L+n,y_{\min})$, and $\lambda(c)=q_0$ everywhere else.
\end{proposition}
\begin{proof}
Let $t_0\ge 1$ be a round from which $P_t=P_\infty$ for all $t\ge t_0$. For
$c\in P_\infty$ we have $c\in P_t\cap P_{t-1}$ for $t>t_0$, so
$\lambda_t(c)=m^+$ by Lemma~\ref{lem:mark5}. The unoccupied cells with an
occupied horizontal neighbor are exactly $(x_L-1,y_{\min})$ and
$(x_L+n,y_{\min})$, because the horizontal neighbors of a cell outside row
$y_{\min}$ also lie outside row $y_{\min}$ and are therefore unoccupied; by
Lemma~\ref{lem:stable} these two cells are at $\ell_2$ from round $t_0+2$ on.
Every other cell is unoccupied and has no occupied horizontal neighbor after
$t_0$, so neither G3 nor, after one round, G4 applies to it, and G5, G6 drive
it to $q_0$ within three rounds, where G6 keeps it. No clause changes any of
these values again, so the configuration is a fixed point.
\end{proof}

The two end cells are the whole difference: under the four-state rules they
alternate forever, here they rest at $\ell_2$. One state buys the fixed
point. We do not claim that five states are necessary;
whether some four-state field is field-silent is open.

\section{Gathering from Arbitrary Configurations with Three States}
\label{sec:gathering}

We now drop connectivity. The robots start on arbitrary distinct cells and
may be so far apart that no robot ever sees another unless the algorithm
brings them together; by Proposition~\ref{prop:inert}(ii) this is hopeless on
an inert floor. Three states suffice:
$\Q=\{q_0,\stw,\sta\}$, where $q_0$ is \emph{blank}, $\stw$ (``west'') means
that some robot stands strictly east of the cell, and $\sta$ (``axis'') marks
a robot's cell and the cells above it in its column.

\subsection{The Rules}

The field transition and the robot algorithm are, as before, given by the
first applicable clause:

\begin{center}\small
\begin{tabular}{clc}
\hline
Rule & Condition on cell $c$ & Next state \\
\hline
H1 & $c$ carries a robot & $\sta$ \\
H2 & $\lambda_\rE\in\{\sta,\stw\}$ or $\lambda_\rN=\stw$ or $\lambda_\rS=\stw$ & $\stw$ \\
H3 & $\lambda_\rS=\sta$ & $\sta$ \\
H4 & otherwise & $\lambda_c$ (unchanged) \\
\hline
\end{tabular}
\qquad
\begin{tabular}{clc}
\hline
Rule & Condition & Action \\
\hline
T1 & $\lambda_\rE\in\{\stw,\sta\}$ & move E \\
T2 & $\lambda_\rS=\sta$ & move S \\
T3 & $\lambda_\rE=q_0\wedge\lambda_\rN=\stw$ & move N \\
T4 & otherwise & stay \\
\hline
\end{tabular}
\end{center}

\begin{figure}[t]
\centering
\setlength{\tabcolsep}{3pt}
\input{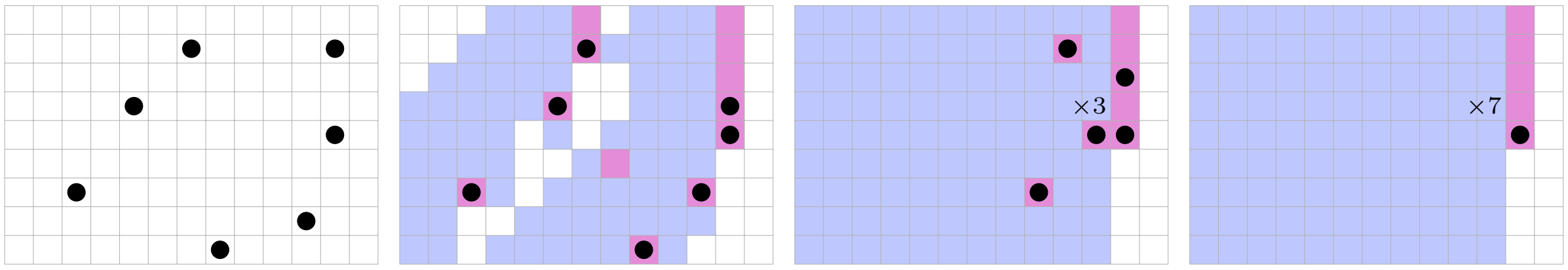}
\caption{Gathering seven scattered robots with three states. Blue cells are
$\stw$, magenta cells are $\sta$ (a robot's cell is always $\sta$), white
cells are blank; ``$\times k$'' counts the robots on a cell. The axis
grows upward from every robot (H3) and the west region floods everything west
of the easternmost robots (H2), overwriting the axes there. Robots walk east
through painted cells (T1); the robot that starts just west of the easternmost
column but below its lowest robot climbs (T3) until it can enter. All robots
meet on the lowest robot of the easternmost column, here $(11,4)$.}
\label{fig:gathering}
\end{figure}

The idea is to make every robot walk to the easternmost occupied column and
then down that column (Fig.~\ref{fig:gathering}). By H1 and H3 an \emph{axis}
grows upward from every robot. By H2 a cell that has an axis or west cell to
its east becomes a west cell, and the west state also spreads north and south;
hence $\stw$ floods the whole half-plane west of the easternmost robots, while
nothing east of them is ever painted, because no rule lets information travel
east. H2 precedes H3, so west of the easternmost column the axes are
overwritten and only the axis of that column survives. A robot walks east as
long as the cell to its east is painted (T1). It thereby reaches the column
just west of the easternmost one; if the cell to its east is an axis cell it
enters the column and descends the axis (T2) onto the lowest robot there. If
instead it arrives below that robot, the cell to its east is blank and will
remain so; such a robot recognizes its situation from the west cell above it
and climbs (T3) until the cell to its east is the meeting cell.

\subsection{Correctness}

Fix an arbitrary initial configuration $P_0$ and let
\[
  X=\max\{x\mid (x,y)\in P_0\},\qquad
  y^\ast=\min\{y\mid (X,y)\in P_0\},\qquad g=(X,y^\ast).
\]
We show that all robots meet on $g$, the cell of the lowest robot of the
easternmost column.

\begin{lemma}[Nothing happens in the east]\label{lem:east}
At every round: (i) no robot is in a column $x>X$; (ii) every cell of a column
$x>X$ is blank; (iii) no cell of column $X$ is in state $\stw$.
\end{lemma}
\begin{proof}
Induction on $t$; all three hold at $t=0$. Assume them at $t$. A robot in
column $X$ has $\lambda_\rE=q_0$ by (ii), so T1 does not apply, and T2, T3 are
vertical; this gives (i) at $t+1$. A cell $c$ with $x>X$ is unoccupied at
$t+1$, its east, north and south neighbors are blank by (ii), so none of
H1--H3 applies and H4 keeps it blank. A cell of column $X$ can become $\stw$
only by H2; its east neighbor is blank by (ii) and its north and south
neighbors lie in column $X$ and are not $\stw$ by (iii).
\end{proof}

\begin{lemma}[The meeting cell]\label{lem:meet}
At every round: (i) every cell $(X,y)$ with $y<y^\ast$ is blank and
unoccupied; (ii) a robot on $g$ does not move, so $g$ is always occupied; and
(iii) $\lambda_t(X,y^\ast+k)=\sta$ for all $k\ge 0$ and $t\ge k+1$.
\end{lemma}
\begin{proof}
(i) and (ii) by induction on $t$. A robot on $g$ has $\lambda_\rE=q_0$ by
Lemma~\ref{lem:east}, $\lambda_\rS=q_0$ by (i), and $\lambda_\rN\ne\stw$ by
Lemma~\ref{lem:east}(iii), so T4 applies. A cell $(X,y)$ with $y<y^\ast$ can
be entered only from the west by T1, from the north by T2, or from the south
by T3; the first two need the cell to be non-blank and the third needs an
occupied cell below it, contradicting (i) at round $t$. It stays blank because
its east neighbor is blank, its vertical neighbors are not $\stw$
(Lemma~\ref{lem:east}) and its south neighbor is blank by (i).
(iii) By (ii) and H1, $\lambda_t(g)=\sta$ for $t\ge 1$. Inductively, for
$k\ge 1$, at the update of every round $t\ge k$ the cell $(X,y^\ast+k)$ is
occupied, and H1 applies, or its south neighbor is $\sta$, H2 does not apply
by Lemma~\ref{lem:east}, and H3 applies.
\end{proof}

\begin{lemma}[The easternmost column]\label{lem:colX}
A robot in column $X$ never leaves it, moves only south, and reaches $g$ in
finitely many rounds.
\end{lemma}
\begin{proof}
T1 is excluded by Lemma~\ref{lem:east}(ii) and T3 by
Lemma~\ref{lem:east}(iii), since the cell above the robot lies in column $X$.
By Lemma~\ref{lem:meet}(i) the robot is in a row $y\ge y^\ast$, and by
Lemma~\ref{lem:meet}(iii) the cell below it is $\sta$ from round $y-y^\ast$
on if $y>y^\ast$, so T2 applies in every later round until the robot is on
$g$, where it stays.
\end{proof}

\begin{lemma}[Paint]\label{lem:paint}
(i) A non-blank cell never becomes blank. (ii) Every cell $(x,y)$ with $x<X$
and $y\ge y^\ast$ is non-blank from some round on, and whenever it is
unoccupied after that round it is in state $\stw$.
\end{lemma}
\begin{proof}
(i) H1--H3 produce $\sta$ or $\stw$ and H4 preserves the state. (ii) By
induction on $X-x$. The east neighbor of $(x,y)$ is $(X,y)$, which is $\sta$
from round $y-y^\ast+1$ on by Lemma~\ref{lem:meet}(iii), or a cell that is
non-blank from some round on by the induction hypothesis and (i). From then
on, at every update, H1 applies if the cell is occupied and H2 applies
otherwise.
\end{proof}

\begin{theorem}\label{thm:gathering}
The three-state algorithm solves Gathering from arbitrary configurations.
Every execution terminates with all robots on
$g=(X,y^\ast)$, the cell of the lowest robot of the easternmost occupied
column of the initial configuration.
\end{theorem}
\begin{proof}[Proof sketch]
\emph{If the robots stop, they have gathered.} Suppose that from some round
on no robot moves and that some robot is not on $g$. By
Lemma~\ref{lem:colX} every robot of column $X$ is on $g$, so the robots off
$g$ are in columns $x<X$; let $r$, on $(x,y)$, be one with the largest row.
Since T1 never applies to $r$, the cell $(x+1,y)$ is blank forever, hence
$y<y^\ast$ by Lemma~\ref{lem:paint}(ii) and Lemma~\ref{lem:meet}(iii). The
cells $(x,y')$ with $y<y'\le y^\ast$ are unoccupied by the choice of $r$. By
Lemma~\ref{lem:paint}(ii) the cell $(x,y^\ast)$ is eventually $\stw$, and
then H2, which precedes H3, turns $(x,y^\ast-1)$, $(x,y^\ast-2)$, \dots,
$(x,y+1)$ into $\stw$ one after the other, whatever their previous states. So
eventually $\lambda_\rE=q_0$ and $\lambda_\rN=\stw$ at $r$, and T3 applies
unless T2 does; in either case $r$ moves, a contradiction.

\emph{The robots stop.} No robot moves west or leaves the columns $x\le X$,
so there are at most $\sum_r (X-x_r)$ moves east, where $x_r$ is the initial
column of robot $r$. Between two of them a robot moves vertically only by T2,
down an axis, that is, towards a cell that a robot of its own column occupies
or has occupied, or by T3, upward through west cells while the cell to its
east is blank. A cell that a robot leaves upward becomes $\stw$ (H2 reads the
west cell that the robot is entering), so a climbing robot is not pulled back
by T2; a robot that descends onto another robot merges with it, and merged
robots never separate; and a climb ends at the latest in row $y^\ast$, where
the cell to the east is painted by Lemma~\ref{lem:paint}. A complete proof
has to bound the vertical moves made while the paint is still spreading; we
omit this routine but lengthy bookkeeping of the transient states.
\end{proof}

We complemented the proof sketch by machine verification
(Section~\ref{sec:sim}): the algorithm gathers on $g$ from every one of the
$4{,}445$ placements of at most three robots in a $5\times5$ box and of four
robots in a $4\times4$ box, and from every random configuration of up to $100$
robots that we tried; the number of rounds grows linearly with the diameter of
the initial configuration.

\subsection{Field-Silence Is Impossible from Arbitrary Configurations}

The three-state algorithm is terminating but not field-silent: after the
robots have met, the west region keeps spreading west and the axis north,
forever. This is not a defect of the algorithm.

\begin{proposition}\label{prop:nosilence}
No algorithm, with any number of field states, solves Gathering from
arbitrary configurations field-silently. The same holds for Line Formation
from arbitrary configurations.
\end{proposition}
\begin{proof}
Let $(\varphi,\mathcal{F})$ be field-silent and consider the execution
$\mathcal{E}_1$ of a single robot started on the origin. It reaches a fixed
point at some round $t_0$. Since only finitely many cells are non-quiescent at
any round, the set of cells that are occupied or non-quiescent at some round
of $\mathcal{E}_1$ is finite and lies within distance $\varrho$ of the origin,
for some $\varrho$.

Now start two robots on $(0,0)$ and $(L,0)$ with $L\ge 2\varrho+4$. By
induction on $t$, the configuration at round $t$ is the superposition of
$\mathcal{E}_1$ and of its translate by $(L,0)$: a cell within distance
$\varrho+1$ of one of the two origins has all its neighbors at distance at
least $\varrho+2$ from the other origin, so $\delta$ and $\varphi$ receive
exactly the arguments they receive in the corresponding single-robot
execution, and every other cell is quiescent with quiescent neighbors and
stays quiescent because $\delta(q_0,\dots,q_0,0)=q_0$. Hence both robots stop
within distance $\varrho$ of their own origins, at distance at least four from
each other: they have neither met nor formed a line.
\end{proof}

The proof uses only that information travels at most one cell per round and
that a field-silent execution of one robot disturbs a bounded region. For
robots that must find each other across unbounded distances, an unbounded
amount of field activity is the price, whatever the number of states.

\section{Line Formation from Arbitrary Configurations with Six States}
\label{sec:linearb}

Gathering lets the robots pile up on one cell. Line Formation from arbitrary
configurations asks for more: the robots must come together from afar
\emph{without ever colliding} and end on consecutive cells of a row. The
difficulty is local. Robots converge on a common cell from the west and from
the north, the two robots are diagonal neighbors, and neither sees the other.
We resolve it with a \emph{landing zone}: a cell that knows a robot stands
directly above it, and that the two robots read differently.

\subsection{The Rules}

Let $\Q=\{q_0,\ste,\stw,\sta,\str,\stl\}$. From round~$1$ on a cell is in
state $\str$ exactly when it carries a robot. The other non-blank states
describe where robots are
relative to the cell: $\stw$ (``west region'') says that some robot stands
strictly east of the cell, in its row or below; $\sta$ (``axis'') that a robot
stands below the cell in its column; $\ste$ (``east region'') that some robot
stands strictly west of the cell, in its row or below. Finally $\stl$
(``landing zone'') marks a cell of type $\sta$ or $\ste$ whose north neighbor
carries a robot.

\begin{center}\small
\begin{tabular}{clc}
\hline
Rule & Condition on cell $c$ & Next state \\
\hline
K1 & $c$ carries a robot & $\str$ \\
K2 & $\lambda_\rN=\str\ \wedge\ \lambda_\rE\in\{\str,\sta,\stw\}$ & $\stw$ \\
K3 & $\lambda_\rN=\str\ \wedge\ \bigl(\lambda_\rS\in\{\str,\sta\}\ \vee\ \lambda_\rW\in\{\str,\sta,\ste,\stl\}\ \vee\ \lambda_c\in\{\ste,\sta,\stl\}\bigr)$ & $\stl$ \\
K4 & $\lambda_\rE\in\{\str,\sta,\stw\}$ & $\stw$ \\
K5 & $\lambda_\rS\in\{\str,\sta\}$ & $\sta$ \\
K6 & $\lambda_\rW\in\{\str,\sta,\ste,\stl\}$ & $\ste$ \\
K7 & $\lambda_c=\stl$ & $\ste$ \\
K8 & otherwise & $\lambda_c$ (unchanged) \\
\hline
\end{tabular}
\end{center}

\begin{center}\small
\begin{tabular}{clc}
\hline
Rule & Condition & Action \\
\hline
U1 & $\neg\rho_\rS\wedge\lambda_\rS=\stl$ & move S \\
U2 & $\neg\rho_\rE\wedge\lambda_\rE\in\{\stw,\sta\}$ & move E \\
U3 & $\rho_\rS\wedge\neg\rho_\rN\wedge\neg\rho_\rE\wedge\lambda_\rE\ne\stl$ & move E \\
U4 & otherwise & stay \\
\hline
\end{tabular}
\end{center}

\begin{figure}[t]
\centering
\setlength{\tabcolsep}{4pt}
\input{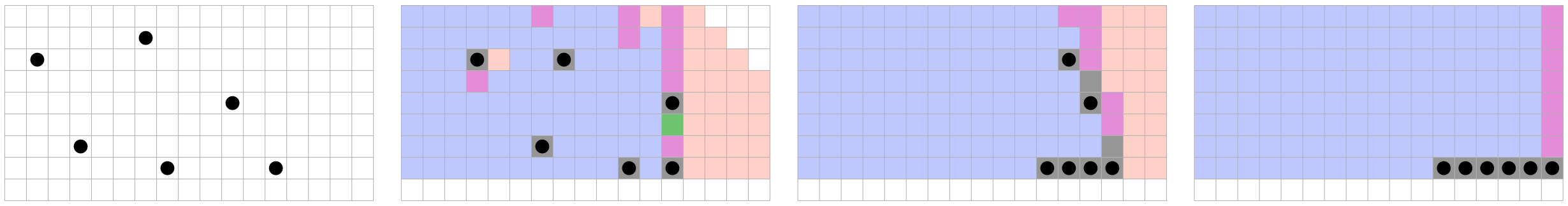}
\caption{Line Formation from an arbitrary configuration with six states, $n=6$.
Blue cells are $\stw$, magenta $\sta$, pink $\ste$, green $\stl$; a robot's
cell is $\str$ (grey). Robots walk east through the west region and the axes
(U2), descend through landing zones (U1), and the top robot of a stack slides
east and drops (U3, U1). The line forms in the lowest initial row.}
\label{fig:linearb}
\end{figure}

K4--K6 paint the three regions around every robot, with the west region
taking precedence over the axis and the axis over the east region. On their
own these regions already bring the robots together: a robot that moves south
onto $\ste$ or $\sta$ and east onto $\stw$ or $\sta$ ends at the easternmost
column of the lowest row. The remaining rules make this approach
collision-free. K2 and K3 concern a cell directly below a robot: if its paint
is the west region it stays $\stw$, since no robot descends into the west
region; if it is an axis or east-region cell it becomes a landing zone. K7
removes the mark once the robot above has left and no neighbor repaints the
cell.

The robot rules use the landing zone as a mutual exclusion device. A robot
descends only into a landing zone (U1), and it never moves east into one (U2,
U3). Because the mark appears one round after the robot above has arrived, a
robot approaching from the west gets in first, and the robot above then finds
the cell occupied. U3 is the rule that turns a pile into a line, as in
Section~\ref{sec:line}: the top robot of a vertical stack slides east off the
stack, and U1 lets it drop beside it (Fig.~\ref{fig:linearb}).

\subsection{Correctness}

Let $y_{\min}$ be the lowest row of $P_0$, as in Section~\ref{sec:line}.

\begin{lemma}[The ground]\label{lem:ground}
At every round, every cell of a row below $y_{\min}$ is blank and unoccupied.
\end{lemma}
\begin{proof}
Induction on $t$. Robots move east and south only, and a robot of row
$y_{\min}$ moves south only by U1, which needs the cell below it to be
$\stl$; so no robot enters the rows below $y_{\min}$ as long as they are
blank. A blank unoccupied cell of row $y_{\min}-1$ stays blank: its east,
west and south neighbors are blank, so K4--K6 and the first two disjuncts of
K3 fail, K2 fails, and $\lambda_c$ is neither $\ste$, $\sta$ nor $\stl$. In
lower rows the north neighbor is blank as well.
\end{proof}

\begin{lemma}[Collision-freedom]\label{lem:collarb}
$|P_t|=n$ for every $t$.
\end{lemma}
\begin{proof}
Robots move east (U2, U3) and south (U1) only, and every rule requires the
target cell to be unoccupied at the $\LL$ of that round. Two robots that enter
the same cell $c$ in the same round therefore come from $c+\rW$ and
$c+\rN$. The robot on $c+\rN$ applies U1, which requires
$\lambda(c)=\stl$; the robot on $c+\rW$ applies U2, which requires
$\lambda(c)\in\{\stw,\sta\}$, or U3, which requires $\lambda(c)\ne\stl$.
Both read $\lambda(c)$ in the same round, a contradiction.
\end{proof}

\begin{lemma}[Finitely many moves]\label{lem:finite}
Every robot moves finitely often.
\end{lemma}
\begin{proof}
By Lemma~\ref{lem:ground} no robot goes below row $y_{\min}$, so there are
finitely many moves south; consider the execution after the last one. Rows
are now fixed, and within a row the robots keep their order, since no robot
enters an occupied cell.

The paint is truthful even when it is stale. By induction on $t$: whenever a
cell $c$ is $\stw$, some robot stands strictly east of $c$, in its row or
below; and whenever $c$ is $\sta$, some robot stands strictly below $c$, in
its column or east of it. Indeed K2 and K4 derive the first fact, and K5 the
second, from the same facts about the east or south neighbor, K8 keeps a
state, and both facts are preserved because robots move east and south only.

We show by induction on the row, from $y_{\min}$ upward, that every robot of
the row moves east finitely often. Let $B$ be the largest column ever reached
by a robot of a lower row, which is finite by the induction hypothesis, and
let $r$ be the easternmost robot of the row. It moves east by U3 only while
it stands on a robot of the row below, hence from a column at most $B$. It
moves east by U2 only into a cell that is $\stw$ or $\sta$; by truthfulness
some robot then stands in a column east of $r$, in the row of $r$ or below,
and since no robot of its own row is east of $r$, that robot belongs to a
lower row, so $r$ is in a column smaller than $B$. Hence $r$ never passes
column $B+1$, and the other robots of the row stay west of $r$. In row
$y_{\min}$ there is no lower row: U3 never applies by Lemma~\ref{lem:ground},
and by truthfulness U2 never applies to $r$.
\end{proof}

\begin{theorem}\label{thm:linearb}
The six-state algorithm solves Line Formation from arbitrary configurations.
Every execution terminates in a line that lies in row $y_{\min}$.
\end{theorem}
\begin{proof}[Proof sketch]
Collision-freedom is Lemma~\ref{lem:collarb} and termination is
Lemma~\ref{lem:finite}. Consider the robots after their last move. From then
on no cell is a landing zone at the $\LL$ of a round: such a cell is
unoccupied and has a robot directly above it, and U1 would move that robot.
The field keeps changing only by propagation, and in every bounded region it
stabilizes to the following pattern, by induction from east to west and from
bottom to top. An unoccupied cell of a row $y\ge y_{\min}$ is $\stw$ if some
robot stands strictly east of it in its row or below; otherwise it is $\sta$
if a robot stands below it in its column; otherwise some robot stands
strictly west of it in its row or below, and it is $\ste$. A cell of the last
two kinds cannot have a robot directly above it, since K3 would turn it into
a landing zone.

Suppose some robot is above row $y_{\min}$, and let $r$, on $(x,k)$, be the
easternmost robot of the highest occupied row. The cells above $r$ and east
of $r$ in its row are unoccupied. If a robot stood below $r$, U3 would be
enabled. So $(x,k-1)$ is unoccupied, and as $r$ stands directly above it, it
is a cell of the first kind: some robot stands strictly east of column $x$ in
a row below $k$. Then $(x+1,k)$, which is unoccupied and has no robot above
it, is $\stw$ or $\sta$, and U2 is enabled, a contradiction.
Hence all robots are in row $y_{\min}$. If two of them were separated by an
unoccupied cell, the western one would see $\stw$ to its east and U2 would be
enabled. So the occupied cells are consecutive.
\end{proof}

In all our experiments the terminal line moreover contains the easternmost
initial column: it extends from that column to the west by as many robots as
arrive along row $y_{\min}$ and to the east by the robots that slide off the
stack. We verified the algorithm on all $5{,}237$ placements of at most three
robots in a $5\times5$ box, four robots in a $4\times4$ box and five robots in
a $3\times4$ box, and on random configurations of up to $30$ robots
(Section~\ref{sec:sim}). The landing zone is essential: the same rules without
it collide in about $30\%$ of random executions. By
Proposition~\ref{prop:nosilence} the algorithm cannot be field-silent, and
indeed its regions spread forever.

\section{Exploration}
\label{sec:exploration}

Our third problem is not a formation problem and has no terminal
configuration. A single robot must visit every cell of a finite arena again
and again. On an inert floor it cannot (Proposition~\ref{prop:inert}(iii)): an
oblivious robot that always sees the same thing always does the same thing.
The floor supplies exactly what is missing, a memory of how the robot left
each cell the last time.

\subsection{Four States: The Floor as a Rotor-Router}

Let $\Q=\{\rN,\rE,\rS,\rW\}$, the state of a cell being read as a pointer,
with $q_0=\rN$, and let $\mathrm{turn}$ be the cyclic shift
$\rN\mapsto\rE\mapsto\rS\mapsto\rW\mapsto\rN$.

\begin{center}\small
\begin{tabular}{clc}
\hline
Rule & Condition on cell $c$ & Next state \\
\hline
X1 & $c$ carries a robot & $\mathrm{turn}(\lambda_c)$ \\
X2 & otherwise & $\lambda_c$ \\
\hline
\end{tabular}
\qquad
\begin{tabular}{clc}
\hline
Rule & Condition & Action \\
\hline
Y & $\lambda_c=d$ & move $d$ \quad ($d\in\Dir$) \\
\hline
\end{tabular}
\end{center}

The robot moves in the direction its own cell points to, and a cell advances
its pointer in every round in which a robot stands on it at the update. The
field never looks at a neighbor and the robot never looks at one either.
On a bounded grid a move towards a missing cell leaves the robot in place;
no special rule is needed for the border.

\begin{lemma}\label{lem:rotor}
Call a round in which the robot stands on cell $v$ at its $\LL$ a
\emph{$v$-round}. In consecutive $v$-rounds the robot attempts consecutive
directions of the cyclic order $\rN,\rE,\rS,\rW$.
\end{lemma}
\begin{proof}
Let $t<t'$ be consecutive $v$-rounds. If the robot is still on $v$ after the
$\MM$ phase of round $t$, because its move was blocked or, on a torus of
width or height one, led back to $v$, then $t'=t+1$ and $v$ was occupied at
the update of round $t$. Otherwise the robot leaves $v$ in round $t$, $v$ is
unoccupied
at every update until the robot re-enters it in the $\MM$ phase of round
$t'-1$, and $v$ is occupied at the update of round $t'-1$. In both cases X1
applies to $v$ exactly once between the two $\LL$ phases.
\end{proof}

\begin{theorem}\label{thm:explore4}
The four-state rotor field solves Exploration of every torus $T_{w,h}$ and of
every bounded grid $B_{w,h}$. On $T_{w,h}$ with $w,h\ge 2$ the cover time is
$wh+h-1$ from every start cell; on $B_{w,h}$ it is $O(wh(w+h))$.
\end{theorem}
\begin{proof}
\emph{Perpetual exploration.} The arena is finite, so some cell $v$ has
infinitely many $v$-rounds. By Lemma~\ref{lem:rotor} every direction is
attempted in one of any four consecutive $v$-rounds, so the robot enters
every existing neighbor of $v$ infinitely often, and the same then holds for
the neighbors of these cells. As the arena is connected, every cell is
visited infinitely often.

\emph{Cover time on a torus.} By symmetry let the robot start on $(0,0)$. In
round $0$ it reads $\rN$ and moves to $(0,1)$. A cell that is entered for the
first time turns from $\rN$ to $\rE$ on arrival, so the robot leaves it
eastward: it runs east along row $1$ and re-enters $(0,1)$ after $w$ rounds.
This cell now turns to $\rS$ and sends the robot to $(0,0)$, whose pointer is
still $\rN$ and turns to $\rE$; the robot runs east along row $0$, re-enters
$(0,0)$, is sent south to $(0,h-1)$, and so on. The rows are thus swept in the
order $1,0,h-1,h-2,\dots,2$; each sweep takes $w$ rounds and is followed by
one round that changes the row, except that the last sweep has seen its whole
row after $w-1$ rounds. With the initial round this gives
$1+(h-1)(w+1)+(w-1)=wh+h-1$ rounds.

\emph{Cover time on a bounded grid.} By Lemma~\ref{lem:rotor}, if the rounds
in which the robot is blocked are deleted, the walk is the rotor-router walk
on the grid graph with the port order induced by $\rN,\rE,\rS,\rW$, and at
most three blocked rounds occur between consecutive steps. A rotor-router walk
covers a graph with $m$ edges and diameter $D$ within $2mD$
steps~\cite{YWB03}; here $m<2wh$ and $D<w+h$.
\end{proof}

After the cover time the walk does not become irregular: a rotor walk on a
finite Eulerian graph locks into an Eulerian circuit~\cite{PDDK96,BGHIKK09,HLMPPW08},
in which every cell is left once in each direction. Counting the blocked
attempts on a bounded grid, one period lasts exactly $4wh$ rounds on both
arenas, which is what we observe in simulation for all sizes we tried
(Section~\ref{sec:sim}). In the long run every cell is therefore revisited
within $4wh$ rounds. Fig.~\ref{fig:exploration} shows the order of first
visits.

\begin{figure}[t]
\centering
\setlength{\tabcolsep}{10pt}
\begin{tabular}{@{}ccc@{}}
\begin{tikzpicture}[x=0.5cm,y=0.5cm,baseline=(current bounding box.south)]
\draw[black!30,line width=0.2pt,step=1] (0,0) grid (6,4);
\node[font=\tiny,inner sep=0pt] at (1.5,1.5) {0};
\node[font=\tiny,inner sep=0pt] at (1.5,2.5) {1};
\node[font=\tiny,inner sep=0pt] at (2.5,2.5) {2};
\node[font=\tiny,inner sep=0pt] at (3.5,2.5) {3};
\node[font=\tiny,inner sep=0pt] at (4.5,2.5) {4};
\node[font=\tiny,inner sep=0pt] at (5.5,2.5) {5};
\node[font=\tiny,inner sep=0pt] at (0.5,2.5) {6};
\node[font=\tiny,inner sep=0pt] at (2.5,1.5) {9};
\node[font=\tiny,inner sep=0pt] at (3.5,1.5) {10};
\node[font=\tiny,inner sep=0pt] at (4.5,1.5) {11};
\node[font=\tiny,inner sep=0pt] at (5.5,1.5) {12};
\node[font=\tiny,inner sep=0pt] at (0.5,1.5) {13};
\node[font=\tiny,inner sep=0pt] at (1.5,0.5) {15};
\node[font=\tiny,inner sep=0pt] at (2.5,0.5) {16};
\node[font=\tiny,inner sep=0pt] at (3.5,0.5) {17};
\node[font=\tiny,inner sep=0pt] at (4.5,0.5) {18};
\node[font=\tiny,inner sep=0pt] at (5.5,0.5) {19};
\node[font=\tiny,inner sep=0pt] at (0.5,0.5) {20};
\node[font=\tiny,inner sep=0pt] at (1.5,3.5) {22};
\node[font=\tiny,inner sep=0pt] at (2.5,3.5) {23};
\node[font=\tiny,inner sep=0pt] at (3.5,3.5) {24};
\node[font=\tiny,inner sep=0pt] at (4.5,3.5) {25};
\node[font=\tiny,inner sep=0pt] at (5.5,3.5) {26};
\node[font=\tiny,inner sep=0pt] at (0.5,3.5) {27};
\end{tikzpicture} &
\begin{tikzpicture}[x=0.5cm,y=0.5cm,baseline=(current bounding box.south)]
\draw[black!30,line width=0.2pt,step=1] (0,0) grid (6,4);
\node[font=\tiny,inner sep=0pt] at (1.5,1.5) {0};
\node[font=\tiny,inner sep=0pt] at (1.5,2.5) {1};
\node[font=\tiny,inner sep=0pt] at (2.5,2.5) {2};
\node[font=\tiny,inner sep=0pt] at (3.5,2.5) {3};
\node[font=\tiny,inner sep=0pt] at (4.5,2.5) {4};
\node[font=\tiny,inner sep=0pt] at (5.5,2.5) {5};
\node[font=\tiny,inner sep=0pt] at (0.5,2.5) {6};
\node[font=\tiny,inner sep=0pt] at (1.5,3.5) {8};
\node[font=\tiny,inner sep=0pt] at (2.5,3.5) {9};
\node[font=\tiny,inner sep=0pt] at (3.5,3.5) {10};
\node[font=\tiny,inner sep=0pt] at (4.5,3.5) {11};
\node[font=\tiny,inner sep=0pt] at (5.5,3.5) {12};
\node[font=\tiny,inner sep=0pt] at (0.5,3.5) {13};
\node[font=\tiny,inner sep=0pt] at (1.5,0.5) {15};
\node[font=\tiny,inner sep=0pt] at (2.5,0.5) {16};
\node[font=\tiny,inner sep=0pt] at (3.5,0.5) {17};
\node[font=\tiny,inner sep=0pt] at (4.5,0.5) {18};
\node[font=\tiny,inner sep=0pt] at (5.5,0.5) {19};
\node[font=\tiny,inner sep=0pt] at (0.5,0.5) {20};
\node[font=\tiny,inner sep=0pt] at (2.5,1.5) {23};
\node[font=\tiny,inner sep=0pt] at (3.5,1.5) {24};
\node[font=\tiny,inner sep=0pt] at (4.5,1.5) {25};
\node[font=\tiny,inner sep=0pt] at (5.5,1.5) {26};
\node[font=\tiny,inner sep=0pt] at (0.5,1.5) {27};
\end{tikzpicture} &
\begin{tikzpicture}[x=0.5cm,y=0.5cm,baseline=(current bounding box.south)]
\draw[black!30,line width=0.2pt,step=1] (0,0) grid (6,4);
\node[font=\tiny,inner sep=0pt] at (1.5,1.5) {0};
\node[font=\tiny,inner sep=0pt] at (1.5,2.5) {1};
\node[font=\tiny,inner sep=0pt] at (2.5,2.5) {2};
\node[font=\tiny,inner sep=0pt] at (3.5,2.5) {3};
\node[font=\tiny,inner sep=0pt] at (4.5,2.5) {4};
\node[font=\tiny,inner sep=0pt] at (5.5,2.5) {5};
\node[font=\tiny,inner sep=0pt] at (5.5,1.5) {7};
\node[font=\tiny,inner sep=0pt] at (5.5,0.5) {9};
\node[font=\tiny,inner sep=0pt] at (4.5,0.5) {12};
\node[font=\tiny,inner sep=0pt] at (4.5,1.5) {15};
\node[font=\tiny,inner sep=0pt] at (3.5,0.5) {22};
\node[font=\tiny,inner sep=0pt] at (3.5,1.5) {25};
\node[font=\tiny,inner sep=0pt] at (2.5,0.5) {32};
\node[font=\tiny,inner sep=0pt] at (2.5,1.5) {35};
\node[font=\tiny,inner sep=0pt] at (1.5,0.5) {42};
\node[font=\tiny,inner sep=0pt] at (0.5,0.5) {52};
\node[font=\tiny,inner sep=0pt] at (0.5,1.5) {55};
\node[font=\tiny,inner sep=0pt] at (0.5,2.5) {58};
\node[font=\tiny,inner sep=0pt] at (1.5,3.5) {60};
\node[font=\tiny,inner sep=0pt] at (2.5,3.5) {61};
\node[font=\tiny,inner sep=0pt] at (3.5,3.5) {62};
\node[font=\tiny,inner sep=0pt] at (4.5,3.5) {63};
\node[font=\tiny,inner sep=0pt] at (5.5,3.5) {64};
\node[font=\tiny,inner sep=0pt] at (0.5,3.5) {144};
\end{tikzpicture} \\[2pt]
{\footnotesize torus, four states} & {\footnotesize torus, two states} & {\footnotesize bounded grid, four states}
\end{tabular}
\caption{Round of the first visit of each cell for a robot started on $(1,1)$
of a $6\times4$ arena. On the torus both rotor fields sweep the arena row by
row in $wh+h-1=27$ rounds; on the bounded grid the four-state field needs
$144$ rounds.}
\label{fig:exploration}
\end{figure}

\subsection{Two States on a Torus, and Optimality}

On a torus two of the four directions are enough. Let $\Q=\{q_0,q_1\}$; a
cell flips its state in every round in which a robot stands on it at the
update, and the robot moves north if its cell is $q_0$ and east if it is
$q_1$.

\begin{theorem}\label{thm:explore2}
The two-state rotor field solves Exploration of every torus $T_{w,h}$, with
cover time $wh+h-1$ for $w,h\ge 2$. No field with a single state solves
Exploration of a torus with $w,h\ge 2$, so two states are optimal.
\end{theorem}
\begin{proof}
Lemma~\ref{lem:rotor} holds with the cyclic order $\rN,\rE$, and no move is
ever blocked. Some cell $v$ has infinitely many $v$-rounds, so its northern
and its eastern neighbor are entered infinitely often; since every cell of a
torus can be reached from every other by moving north and east only, every
cell is visited infinitely often. For the cover time, the robot moves from
$(0,0)$ to $(0,1)$, runs east along row $1$, re-enters $(0,1)$, which flips
back to $q_0$ and sends it north to $(0,2)$, and so on; the rows are swept in
the order $1,2,\dots,h-1,0$, with the same count as in
Theorem~\ref{thm:explore4}. The lower bound is
Proposition~\ref{prop:inert}(iii).
\end{proof}

The two-state field does not explore bounded grids: the robot ends in the
north-east corner, where both of its moves are blocked. Whether two or three
states suffice there is open; four do, by Theorem~\ref{thm:explore4}.

\begin{remark}
Lemma~\ref{lem:rotor} and the first part of the proof of
Theorem~\ref{thm:explore4} hold for any number of robots, so $k$ robots also
explore the arena perpetually, and in our simulations faster. Robots that
enter the same cell move together from then on, however, so the parallelism
of multi-agent rotor walks is not guaranteed.
\end{remark}

\begin{remark}
On the infinite grid neither rotor field explores: a robot leaves every cell
it enters for the first time eastward, so after its first step it marches east
forever. Rotor walks on $\mathbb{Z}^2$ are known to escape to infinity from
some initial rotor configurations~\cite{FGLP14}. Whether a finite-state field
lets an oblivious robot visit every cell of $\mathbb{Z}^2$ is open.
\end{remark}

\section{Further Examples}
\label{sec:examples}

The algorithms above use the floor as a trace, as a map of regions and as a
memory of departures. This section sketches three more rule sets that use it
in other ways. We do \emph{not} prove them correct; they were validated by
simulation only (Section~\ref{sec:sim}) and are meant as evidence of the range
of the model and as a source of problems.

\subsection{The Floor as a Ruler: Equal Spacing}

Robots stand on distinct cells of a row and must end exactly $k$ cells apart,
for a fixed $k\ge 2$, measured from the westmost robot, which does not move.
Use $k+2$ states
$\{q_0,\str,c_1,\dots,c_{k-1},\mathsf{f}\}$. A robot's cell is $\str$, and the
cells east of a robot count their distance to it: the cell east of $\str$
becomes $c_1$, the cell east of $c_j$ becomes $c_{j+1}$, the cell east of
$c_{k-1}$ or $\mathsf{f}$ becomes $\mathsf{f}$ (``far''), and a cell with
nothing to its west stays blank. A robot only reads the cells beside it:

\begin{center}\small
\begin{tabular}{lll}
\hline
Condition & Action & Meaning \\
\hline
$\lambda_\rW\in\{\str,c_1,\dots,c_{k-2}\}\ \wedge\ \lambda_\rE=c_1$ & move E & closer than $k$ to the robot on the west \\
$\lambda_\rW=\mathsf{f}$ & move W & farther than $k$ \\
otherwise & stay & $\lambda_\rW=c_{k-1}$: exactly $k$; $\lambda_\rW=q_0$: westmost \\
\hline
\end{tabular}
\end{center}

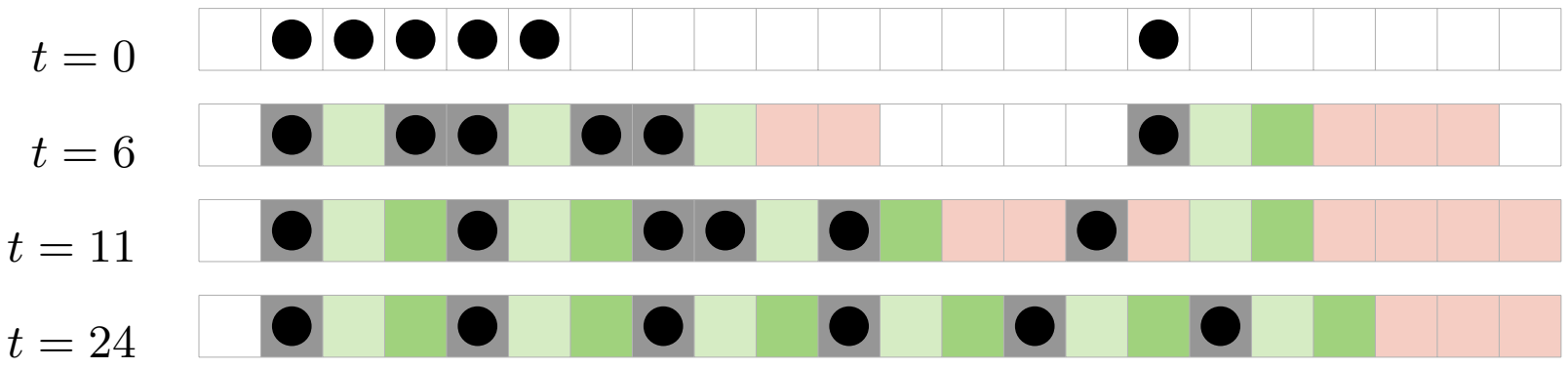
\begin{figure}[t]
\centering
\begin{tabular}{@{}rl@{}}
{\footnotesize $t=0$} & \begin{tikzpicture}[x=0.42cm,y=0.42cm,baseline=(current bounding box.south)]
\draw[black!30,line width=0.2pt,step=1] (0,0) grid (22,1);
\fill[black] (1.5,0.5) circle (0.32);
\fill[black] (2.5,0.5) circle (0.32);
\fill[black] (3.5,0.5) circle (0.32);
\fill[black] (4.5,0.5) circle (0.32);
\fill[black] (5.5,0.5) circle (0.32);
\fill[black] (15.5,0.5) circle (0.32);
\end{tikzpicture} \\[3pt]
{\footnotesize $t=6$} & \begin{tikzpicture}[x=0.42cm,y=0.42cm,baseline=(current bounding box.south)]
\fill[cabR] (1,0) rectangle ++(1,1);
\fill[cabCa] (2,0) rectangle ++(1,1);
\fill[cabR] (3,0) rectangle ++(1,1);
\fill[cabR] (4,0) rectangle ++(1,1);
\fill[cabCa] (5,0) rectangle ++(1,1);
\fill[cabR] (6,0) rectangle ++(1,1);
\fill[cabR] (7,0) rectangle ++(1,1);
\fill[cabCa] (8,0) rectangle ++(1,1);
\fill[cabF] (9,0) rectangle ++(1,1);
\fill[cabF] (10,0) rectangle ++(1,1);
\fill[cabR] (15,0) rectangle ++(1,1);
\fill[cabCa] (16,0) rectangle ++(1,1);
\fill[cabCb] (17,0) rectangle ++(1,1);
\fill[cabF] (18,0) rectangle ++(1,1);
\fill[cabF] (19,0) rectangle ++(1,1);
\fill[cabF] (20,0) rectangle ++(1,1);
\draw[black!30,line width=0.2pt,step=1] (0,0) grid (22,1);
\fill[black] (1.5,0.5) circle (0.32);
\fill[black] (3.5,0.5) circle (0.32);
\fill[black] (4.5,0.5) circle (0.32);
\fill[black] (6.5,0.5) circle (0.32);
\fill[black] (7.5,0.5) circle (0.32);
\fill[black] (15.5,0.5) circle (0.32);
\end{tikzpicture} \\[3pt]
{\footnotesize $t=11$} & \begin{tikzpicture}[x=0.42cm,y=0.42cm,baseline=(current bounding box.south)]
\fill[cabR] (1,0) rectangle ++(1,1);
\fill[cabCa] (2,0) rectangle ++(1,1);
\fill[cabCb] (3,0) rectangle ++(1,1);
\fill[cabR] (4,0) rectangle ++(1,1);
\fill[cabCa] (5,0) rectangle ++(1,1);
\fill[cabCb] (6,0) rectangle ++(1,1);
\fill[cabR] (7,0) rectangle ++(1,1);
\fill[cabR] (8,0) rectangle ++(1,1);
\fill[cabCa] (9,0) rectangle ++(1,1);
\fill[cabR] (10,0) rectangle ++(1,1);
\fill[cabCb] (11,0) rectangle ++(1,1);
\fill[cabF] (12,0) rectangle ++(1,1);
\fill[cabF] (13,0) rectangle ++(1,1);
\fill[cabR] (14,0) rectangle ++(1,1);
\fill[cabF] (15,0) rectangle ++(1,1);
\fill[cabCa] (16,0) rectangle ++(1,1);
\fill[cabCb] (17,0) rectangle ++(1,1);
\fill[cabF] (18,0) rectangle ++(1,1);
\fill[cabF] (19,0) rectangle ++(1,1);
\fill[cabF] (20,0) rectangle ++(1,1);
\fill[cabF] (21,0) rectangle ++(1,1);
\draw[black!30,line width=0.2pt,step=1] (0,0) grid (22,1);
\fill[black] (1.5,0.5) circle (0.32);
\fill[black] (4.5,0.5) circle (0.32);
\fill[black] (7.5,0.5) circle (0.32);
\fill[black] (8.5,0.5) circle (0.32);
\fill[black] (10.5,0.5) circle (0.32);
\fill[black] (14.5,0.5) circle (0.32);
\end{tikzpicture} \\[3pt]
{\footnotesize $t=24$} & \begin{tikzpicture}[x=0.42cm,y=0.42cm,baseline=(current bounding box.south)]
\fill[cabR] (1,0) rectangle ++(1,1);
\fill[cabCa] (2,0) rectangle ++(1,1);
\fill[cabCb] (3,0) rectangle ++(1,1);
\fill[cabR] (4,0) rectangle ++(1,1);
\fill[cabCa] (5,0) rectangle ++(1,1);
\fill[cabCb] (6,0) rectangle ++(1,1);
\fill[cabR] (7,0) rectangle ++(1,1);
\fill[cabCa] (8,0) rectangle ++(1,1);
\fill[cabCb] (9,0) rectangle ++(1,1);
\fill[cabR] (10,0) rectangle ++(1,1);
\fill[cabCa] (11,0) rectangle ++(1,1);
\fill[cabCb] (12,0) rectangle ++(1,1);
\fill[cabR] (13,0) rectangle ++(1,1);
\fill[cabCa] (14,0) rectangle ++(1,1);
\fill[cabCb] (15,0) rectangle ++(1,1);
\fill[cabR] (16,0) rectangle ++(1,1);
\fill[cabCa] (17,0) rectangle ++(1,1);
\fill[cabCb] (18,0) rectangle ++(1,1);
\fill[cabF] (19,0) rectangle ++(1,1);
\fill[cabF] (20,0) rectangle ++(1,1);
\fill[cabF] (21,0) rectangle ++(1,1);
\draw[black!30,line width=0.2pt,step=1] (0,0) grid (22,1);
\fill[black] (1.5,0.5) circle (0.32);
\fill[black] (4.5,0.5) circle (0.32);
\fill[black] (7.5,0.5) circle (0.32);
\fill[black] (10.5,0.5) circle (0.32);
\fill[black] (13.5,0.5) circle (0.32);
\fill[black] (16.5,0.5) circle (0.32);
\end{tikzpicture}
\end{tabular}
\caption{Equal spacing with gap $k=3$ (five states). Green cells count the
distance to the robot on the west ($c_1$ light, $c_2$ dark), orange cells are
far from it. The dense group spreads east and the straggler comes west.}
\label{fig:spacing}
\end{figure}

A robot steps east only into a cell that already shows $c_1$, that is, a cell
that has registered it, and west only into a cell that shows $\mathsf{f}$;
since a cell cannot show both, two robots never enter the same cell from
opposite sides, which is the device of Lemma~\ref{lem:coll} once more. In our
simulations all $1{,}200$ random rows with $k=2,\dots,5$ and up to $12$
robots ended with the robots exactly $k$ apart (Fig.~\ref{fig:spacing}); rows
are independent, so every row of a two-dimensional configuration becomes a
ruler of its own. Uniform spreading on a line is a classical problem for
robots that see their neighbors~\cite{CP08}; here the robots see nothing
beyond the adjacent cells and the floor measures the distance for them.

\subsection{The Floor as an Excitable Medium: Filling a Painted Pattern}

In the problems so far the floor starts blank. A floor can also carry the
\emph{input}. Suppose some cells are painted as \emph{targets} before the
robots start, as a pattern would be painted on a warehouse floor, and the
robots, scattered arbitrarily, must end with exactly one robot on each target.
We allow the initial field to be non-uniform for this purpose and use eight
states: the target states $\mathsf{T}$ (empty target) and $\mathsf{D}$
(target with a robot on it), an inert state for walls, and five states
$q_0,\mathsf{r},w_1,w_2,w_3$ that make the rest of the floor an excitable
medium in the sense of Greenberg and Hastings~\cite{GH78}: a resting cell
($q_0$, never excited, or $\mathsf{r}$, resting again) that has an excited
neighbor ($w_1$) or an empty target next to it becomes excited, and an
excited cell passes through $w_2$ and $w_3$ back to rest. Every empty target
thus emits waves of period four, waves from different targets annihilate
where they meet, and a target stops emitting as soon as a robot stands on it,
so no stale information is left behind.

\begin{figure}[t]
\centering
\setlength{\tabcolsep}{4pt}
\input{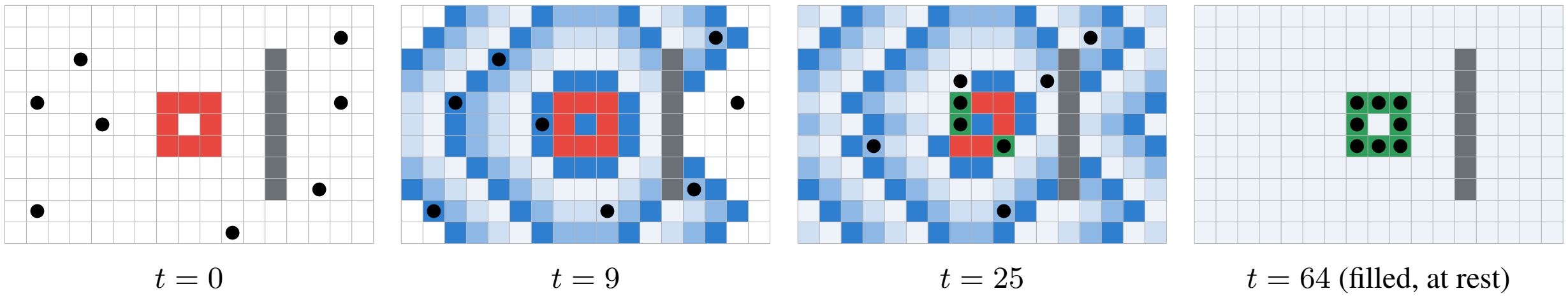}
\caption{Filling a pattern painted on the floor. Red cells are empty targets,
green cells filled ones, the dark column is a wall; blue shades are the phases
of the waves emitted by the empty targets ($w_1$ darkest). Robots follow the
waves, pass around the wall, and the floor comes to rest when every target
is taken.}
\label{fig:filling}
\end{figure}

A robot steps to a neighbor whose phase is one ahead of the phase of its own
cell, that is, to a neighbor that the wave reached one round earlier and that
is therefore closer to the nearest empty target, or onto an adjacent empty
target; a robot on a target stays. The phase is at the same time the traffic
rule. A robot on an excited cell may only move north, on $w_2$ only east, on
$w_3$ only south, and on a resting cell only west. Two robots that enter the
same cell must both see it one phase ahead of their own cells, hence have the
same phase, hence move in the same direction, which is impossible for
distinct robots; the cells around an empty target fire together, so the same
holds for two robots that enter a target. The phase of the wave thus acts as
a clock that multiplexes the four directions. In all our runs with at least
as many robots as targets, every target ended with exactly one robot and the
floor came to rest (Fig.~\ref{fig:filling}); surplus robots simply stopped.

In simulation, $360$ random patterns of up to $20$ targets were filled
without a collision, with as many robots as targets, with a surplus, and with
too few robots (in which case all robots end on targets). With walls the same
rules guide robots through a maze, along shortest paths to the nearest target
that is still empty, because the waves run through the corridors. One
limitation is geometric rather than algorithmic: a target in a corridor of
width one blocks the passage once it is occupied.

\subsection{Simulator and Validation}
\label{sec:sim}

All rule sets of this paper were implemented in a simulator of the model, in
which a rule set is a short list of guarded clauses for $\delta$ and for
$\varphi$, exactly as in the tables above. Table~\ref{tab:validation} lists
the instances on which each algorithm was run; every run ended in the
configuration claimed, without a collision where collisions are forbidden.
The exhaustive enumeration for the two algorithms for connected
configurations was run with a separate, direct implementation of their rule
tables; for the five-state algorithm it also checks that the field reaches a
fixed point. These runs complement the proof sketches of
Theorems~\ref{thm:four}, \ref{thm:five}, \ref{thm:gathering}
and~\ref{thm:linearb}; for the rule sets of this section they are the only
evidence.

\begin{table}[t]
\centering
\caption{Instances checked by simulation.}
\label{tab:validation}
\small
\begin{tabular}{lcl}
\toprule
Algorithm & $|\Q|$ & Instances \\
\midrule
Line Formation, connected (Sec.~\ref{sec:line}, \ref{sec:silent}) & 4, 5 & all 691{,}277 connected configurations, $n\le 12$ \\
Gathering, arbitrary (Sec.~\ref{sec:gathering}) & 3 & all 4{,}445 small placements, $n\le4$; 890 random, $n\le 100$ \\
Line Formation, arbitrary (Sec.~\ref{sec:linearb}) & 6 & all 5{,}237 small placements, $n\le5$; 400 random, $n\le 30$ \\
Exploration (Sec.~\ref{sec:exploration}) & 4 & tori and bounded grids from $1\times1$ to $31\times17$ \\
Exploration (Sec.~\ref{sec:exploration}) & 2 & tori from $1\times1$ to $31\times17$ \\
Equal spacing, $k=2,\dots,5$ & $k+2$ & 300 random rows for each $k$ \\
Pattern filling and maze & 8 & 360 random patterns without walls; one maze with four targets \\
\bottomrule
\end{tabular}
\end{table}

\section{Conclusion}
\label{sec:conclusion}

We proposed \CABot, a model in which the plane underlying an LCM system is
an anonymous, uniform cellular automaton and $\LL$ returns cell states along
with robot positions. No capability is added to the robots: they stay
oblivious, silent, lightless and myopic with visibility range~$1$, and cannot
even write to the field. A handful of field states then solves problems that
are out of reach on an inert plane: Line Formation from connected
configurations with four states and field-silently with five, Gathering and
Line Formation from arbitrary configurations with three and six states, and
perpetual Exploration of finite grids with four states and of tori with two.
The floor plays a different part each time: a decaying trace, a map of
regions, a mutual exclusion device, a rotor-router, a ruler, an excitable
medium.

Several directions are open.

\emph{Lower bounds.} Only one of our state counts is known to be optimal,
the two states for exploring a torus. Is Gathering from arbitrary
configurations possible with two states? Is Line Formation from connected
configurations possible with three, or field-silently with four? How many
states does a bounded grid need for Exploration? Even separating two
consecutive values would say something about what a field can store.

\emph{Field-silence.} Proposition~\ref{prop:nosilence} rules out silence
whenever robots must find each other across unbounded distances. A
characterization of the problems that admit field-silent solutions, and of
the price of silence in states, would be the analogue for \CABot\ of the
memory lower bounds for silent stabilization~\cite{DGS99}.

\emph{Schedulers.} \CABot\ is defined here for $\FSYNC$ only. Extending it
to $\SSYNC$ and $\ASYNC$ robots is first of all a modeling question, because
the $\UU$ phase belongs to the field and not to the robots: a scheduler for
\CABot\ must say when the cells update relative to the $\LL$, $\CC$ and $\MM$
phases of the individual robots, and whether the field keeps its own
synchronous clock, follows the activations of the robots, or is itself
subject to an adversary. Only once such schedulers are fixed can one ask
which of our algorithms, whose correctness rests on the timing of a round,
survive them, and at what cost in states. The answer decides how close the
model is to the warehouse floors that motivated it.

\emph{Proofs and the infinite grid.} The timing arguments that we left as
sketches in Theorems~\ref{thm:four}, \ref{thm:five}, \ref{thm:gathering}
and~\ref{thm:linearb} remain to be written out in full, the examples of
Section~\ref{sec:examples} call for proofs, and Exploration of
$\mathbb{Z}^2$ by an oblivious robot on a finite-state floor is open.

\section*{Acknowledgement}
This work was supported by JSPS KAKENHI Grant Numbers JP23K24825, JP25K03078, JP25K03079, JP26K02865, JP26K23809, and MEXT Supporting Pioneering Research through AI for 1,000 Discovery challenges Program (SPReAD) Japan Grant Number JPMXP1726306000.

\end{document}